\documentclass{LMCS}

\usepackage{amsmath,amssymb,amsthm}
\usepackage[T1]{fontenc}
\usepackage{lmodern}
\usepackage{mathpartir}
\usepackage{stmaryrd,hyperref}
\usepackage{helvet,cclicenses}

\theoremstyle{plain}
\newtheorem{theorem}[thm]{Theorem}
\newtheorem{corollary}[thm]{Corollary}
\newtheorem{lemma}[thm]{Lemma}
\newtheorem{proposition}[thm]{Proposition}

\theoremstyle{definition}
\newtheorem{definition}[thm]{Definition}
\newtheorem{example}[thm]{Example}

\newtheorem*{note*}{Note}
\newtheorem*{notes*}{Notes}
\newtheorem*{null*}{}
\newtheorem{remark}[thm]{Remark}

\newcommand{\act}{\mathbin{\mbox{\boldmath$\cdot$}}}
\newcommand{\aeq}{=_{\alpha}}
\newcommand{\ALT}{\mathrel{\kw{|}}}
\newcommand{\AND}{\kw{and}}
\newcommand{\ar}[1][\sigma]{#1}
\newcommand{\arity}{\mathit{arity}}
\newcommand{\Arity}{\mathit{Arity}}

\newcommand{\atm}[1][a]{#1}
\newcommand{\ATM}{\kw{atm}}
\newcommand{\Atom}{\mathbb{A}}
\newcommand{\atoms}{\mathit{atom}}
\newcommand{\bij}{\cong}
\newcommand{\bimp}{\Leftrightarrow}
\newcommand{\BIND}{\kw{bind}}
\newcommand{\BINDTY}{\kw{bnd}}
\newcommand{\BINDVAL}[2]{\mathopen{\text{\normalfont\guillemotleft}}#1
  \mathclose{\text{\normalfont\guillemotright}}#2}
\newcommand{\BOOL}{\kw{bool}}

\newcommand{\comp}{\circ}
\newcommand{\CON}[1][C]{\kw{#1}}
\newcommand{\Con}{\mathcal{C}}
\newcommand{\config}[3]{\langle#1, #2, #3\rangle}
\renewcommand{\conj}{\wedge}
\newcommand{\CR}[1]{\mathrel{\widehat{#1}}}
\newcommand{\defeq}{\triangleq}
\newcommand{\den}[1]{\llbracket#1\rrbracket}
\newcommand{\DIVERGE}{\kw{diverge}}
\newcommand{\dom}{\mathit{dom}}
\newcommand{\dty}{\delta}
\newcommand{\Dty}{\mathcal{D}}
\newcommand{\ELSE}{\mathrel{\kw{else}}}
\newcommand{\ent}{\vdash}
\newcommand{\enty}{\vdash}
\newcommand{\EQ}{\mathbin{\kw{=}}}
\newcommand{\er}{\mathrel{\mathcal{E}}}
\newcommand{\es}{\mathit{Id}}
\renewcommand{\exp}[1][e]{#1}
\newcommand{\Exp}{\mathit{Exp}}
\newcommand{\FALSE}{\kw{false}}
\newcommand{\fmap}{\stackrel{\mathrm{fin}}{\rightarrow}}
\newcommand{\FRESH}{\kw{fresh}}
\newcommand{\FST}{\kw{fst}}
\newcommand{\fun}{\rightarrow}
\newcommand{\FUN}{\kw{fun}}
\newcommand{\FUNTY}{\mathbin{\rightarrow}}
\newcommand{\fv}{\mathit{fv}}
\newcommand{\IF}{\mathop{\kw{if}}}
\newcommand{\imp}{\Rightarrow}
\newcommand{\IN}{\mathrel{\kw{in}}}

\newcommand{\LET}{\mathop{\kw{let}}}
\newcommand{\LP}{\mathopen{\kw{(}}}
\newcommand{\kw}[1]{\mathsf{#1}}
\newcommand{\MATCH}{\kw{match}}
\newcommand{\NAT}{\kw{nat}}
\newcommand{\NNO}{\mathbb{N}}
\newcommand{\Obs}{\mathcal{O}}
\newcommand{\OBS}[1][obs]{\kw{#1}}
\newcommand{\OF}{\mathbin{\kw{of}}}
\newcommand{\ofty}{:}
\newcommand{\OFTY}{\mathrel{\kw{:}}}
\newcommand{\opeq}{\cong}
\newcommand{\opeqo}{\cong^{\circ}}
\newcommand{\ords}{\olessthan}
\newcommand{\PAIR}[2]{\LP#1\mathbin{\kw{,}}#2\RP}
\newcommand{\Perm}{\mathbb{P}}
\newcommand{\PRODTY}{\mathbin{\kw{*}}}
\newcommand{\PROJ}{\kw{proj}}
\newcommand{\rename}[2]{\{#2/#1\}}
\newcommand{\rep}[1]{\ulcorner#1\urcorner}
\newcommand{\RP}{\mathclose{\kw{)}}}
\newcommand{\SND}{\kw{snd}}
\newcommand{\State}{\mathit{State}}
\newcommand{\s}[1][a]{\vec{#1}}
\newcommand{\stk}[1][S]{#1}
\newcommand{\Stk}{\mathit{Stk}}
\newcommand{\sub}[2]{{[#2/#1]}}
\newcommand{\supp}{\mathit{supp}}
\newcommand{\swap}[2]{(#1\;#2)}
\newcommand{\SWAP}{\kw{swap}}
\newcommand{\TERM}{\kw{term}}
\newcommand{\terminates}[1][]{{\downarrow_{#1}}}
\newcommand{\THEN}{\mathrel{\kw{then}}}
\newcommand{\TO}{\mathbin{\rightarrow}}
\newcommand{\trans}{\longrightarrow}
\newcommand{\TRUE}{\kw{true}}
\newcommand{\ty}{\tau}
\newcommand{\Ty}{\mathit{Typ}}
\newcommand{\TYPE}{\kw{type}}
\newcommand{\TYVAR}{\alpha}
\newcommand{\UNBIND}{\kw{unbind}}
\newcommand{\UNIT}{\kw{unit}}
\newcommand{\UNITVAL}{\LP\RP}
\newcommand{\val}[1][v]{#1}
\newcommand{\Val}{\mathit{Val}}
\newcommand{\VAL}{\kw{val}}
\newcommand{\vid}[1][x]{#1}
\newcommand{\Vid}{\mathbb{V}}
\newcommand{\w}[1][w]{#1}
\newcommand{\WITH}{\mathbin{\kw{with}}}
\newcommand{\World}{\mathit{World}}

\def\doi{4 (1:4) 2008}
\lmcsheading%
{\doi}
{1--33}
{}
{}
{Apr.~19, 2007}
{Mar.~18, 2008}
{}   

\begin{document}

\title[Generative Unbinding of Names]{Generative Unbinding of
Names\rsuper *}

\author[A.M.~ Pitts]{Andrew M.~Pitts\rsuper a}	
\address{{\lsuper a}University of Cambridge Computer Laboratory\\
Cambridge CB3 0FD, UK}	
\email{Andrew.Pitts@cl.cam.ac.uk}  
\thanks{{\lsuper a}Research supported by UK~EPSRC grant EP/D000459/1.}	

\author[M.R.~Shinwell]{Mark R.~Shinwell\rsuper b}	
\address{{\lsuper b}CodeSourcery, Ltd}	
\email{mark@three-tuns.net}  




\titlecomment{{\lsuper *}This paper is a revised and expanded version of
  \cite{PittsAM:genun}.}


\begin{abstract}
  \noindent This paper is concerned with the form of typed name
  binding used by the FreshML family of languages. Its characteristic
  feature is that a name binding is represented by an abstract
  (name,value)-pair that may only be deconstructed via the generation
  of fresh bound names.  The paper proves a new result about what
  operations on names can co-exist with this construct.  In FreshML
  the only observation one can make of names is to test whether or not
  they are equal.  This restricted amount of observation was thought
  necessary to ensure that there is no observable difference between
  alpha-equivalent name binders.  Yet from an algorithmic point of
  view it would be desirable to allow other operations and relations
  on names, such as a total ordering.  This paper shows that, contrary
  to expectations, one may add not just ordering, but almost any
  relation or numerical function on names without disturbing the
  fundamental correctness result about this form of typed name binding
  (that object-level alpha-equivalence precisely corresponds to
  contextual equivalence at the programming meta-level), so long as
  one takes the state of dynamically created names into account.
\end{abstract}

\maketitle

\section{Introduction}

FreshML and the language systems that it has inspired provide some
user-friendly facilities within the context of strongly typed
functional programming for computing with syntactical data structures
involving names and name binding.  The underlying theory was presented
in~\cite{PittsAM:metpbn,PittsAM:frepbm} and has been realised in the
Fresh patch of Objective Caml~\cite{ShinwellMR:freona}. FreshML has
also inspired Pottier's C$\alpha$ml tool~\cite{PottierF:ovec} for
Objective Caml and Cheney's FreshLib library~\cite{CheneyJ:scryn} for
Haskell. The approach taken to binding in all these works is
``nominal'' in that the user is given access to the names of bound
entities and can write syntax manipulating programs that follow the
informal practice of referring to $\alpha$-equivalence classes of
terms via representatives.  However, in FreshML the means of access to
bound names is carefully controlled by the type system.  It has been
shown~\cite{ShinwellMR:freafp,PittsAM:monsf} that its static and
dynamic properties combine to guarantee a certain ``correctness of
representation'' property: data structures representing
$\alpha$-equivalent syntactical terms (that is, ones differing only in
the names of bound entities) always behave the same in any program. So
even though programs can name names, as it were, $\alpha$-equivalence
of name bindings is taken care of automatically by the programming
language design.

\begin{figure}
  \centering
  \[
  \begin{array}{ll}
    \TYPE &\ATM\\
    \TYPE &\TYVAR\,\BINDTY\\
    \VAL  &\FRESH\OFTY\UNIT\FUNTY\ATM\\
    \VAL  &\BIND\OFTY\ATM\PRODTY\TYVAR\FUNTY\TYVAR\,\BINDTY\\
    \VAL  &\UNBIND\OFTY\TYVAR\,\BINDTY\FUNTY\ATM\PRODTY\TYVAR\\
    \VAL  &\LP\EQ\RP\OFTY\ATM\FUNTY\ATM\FUNTY\BOOL
  \end{array}
  \]
  \caption{A signature for name binding.}
  \label{fig:signb}
\end{figure}

Of course such a correctness of representation property depends rather
delicately upon which operations on bound names are allowed. At the
heart of this approach to binding is an operation that we call
\emph{generative unbinding}.  To explain what it involves, consider a
simplified version of Fresh Objective Caml with a single type $\ATM$
of bindable names and a parametric family of types $\TYVAR\,\BINDTY$
classifying abstractions of single names over values of type $\TYVAR$.
To explain: both $\ATM$ and $\TYVAR\,\BINDTY$ are abstract types that
come with the signature of operations shown in Figure~\ref{fig:signb}.
The closed values of type $\ATM$ are drawn from a countably infinite
set $\Atom$ of symbols that we call \emph{atoms}. Programs only get
access to atoms by evaluating the expression $\FRESH\UNITVAL$ to get a
fresh one; and hence program execution depends upon a state recording
the atoms that have been created so far. Given a type $\ty$, closed
values of type $\ty\,\BINDTY$ are called \emph{atom bindings} and are
given by pairs $\BINDVAL{\atm}{\val}$ consisting of an atom
$\atm\OFTY\ATM$ and a closed value $\val\OFTY\ty$.  Atom bindings are
constructed by evaluating $\BIND\PAIR{\atm}{\val}$. Fresh Objective
Caml provides a very convenient form of generative pattern-matching
for deconstructing atom bindings. To keep things simple, here we will
avoid the use of pattern-matching and consider an equivalent mechanism
for deconstructing atom binding via an $\UNBIND$ function carrying out
generative unbinding: $\UNBIND\,\BINDVAL{\atm}{\val}$ evaluates by
first evaluating $\FRESH\UNITVAL$ to obtain a fresh atom $\atm'$ and
then returning the pair $\PAIR{\atm'}{\val\rename{\atm}{\atm'}}$,
where in general $\val\rename{\atm}{\atm'}$ denotes the value obtained
from $\val$ by renaming all occurrences of $\atm$ to be $\atm'$. The
instance of renaming that arises when evaluating
$\UNBIND\,\BINDVAL{\atm}{\val}$ is special: the fresh atom $\atm'$
does not occur in $\val$ and so $\val\rename{\atm}{\atm'}$ is
equivalent to the result of applying to $\val$ the semantically better
behaved operation of \emph{swapping} $\atm$ and $\atm'$. Although
implementing such an atom swapping operation on all types of values is
the main extension that the Fresh patch makes to Objective Caml, we
have not included a $\SWAP\OFTY\ATM\FUNTY\ATM\FUNTY\TYVAR\FUNTY\TYVAR$
operation in the signature of Figure~\ref{fig:signb}. This is because
it is possible for users to define atom swapping themselves for
specific types on a case-by-case basis.  Although this approach has
some limitations, is enough for our purposes here.  (The approach is
more useful in the presence of Haskell-style type
classes---see~\cite{CheneyJ:scryn}.)

The type $\TYVAR\,\BINDTY$ is used in data type declarations in the
argument type of value constructors representing binders. To take a
familiar example, the terms of the untyped $\lambda$-calculus (all
terms, whether open or closed, with variables given by atoms
$\atm\in\Atom$)
\[
t ::= \atm \mid \lambda\atm. t \mid t\,t
\]
can be represented by closed values of the type $\TERM$
given by the declaration
\begin{equation}
  \label{eq:10}
  \TYPE
  \begin{array}[t]{rcl}
    \TERM & \EQ  & \CON[V] \OF \ATM\\
    & \ALT & \CON[L] \OF \TERM\,\BINDTY\\
    & \ALT & \CON[A] \OF \TERM\PRODTY\TERM\;.
  \end{array}  
\end{equation}
The value $\rep{t}\OFTY\TERM$ representing a $\lambda$-term $t$ is
defined by
\begin{equation}
  \label{eq:11}
  \begin{array}[c]{rcl}
    \rep{\atm} &\defeq& \CON[V]\,\atm\\
    \rep{\lambda\atm.t} &\defeq& \CON[L]\,\BINDVAL{\atm}{\rep{t}}\\
    \rep{t_1\,t_2} &\defeq& \CON[A]\PAIR{\rep{t_1}}{\rep{t_2}}
  \end{array}
\end{equation}
and satisfies:
\begin{quote}
  \textbf{Correctness of Representation}: \emph{two $\lambda$-terms
    are $\alpha$-equivalent, $t_1\aeq t_2$, iff $\rep{t_1}$ and
    $\rep{t_2}$ are contextually equivalent closed values of type
    $\TERM$, i.e.~can be used interchangeably in any well-typed Fresh
    Objective Caml program without affecting the observable results of
    program execution.}
\end{quote}
Since it is also the case that every closed value of type $\TERM$ is
of the form $\rep{t}$ for some $\lambda$-term $t$, it follows that
there is a bijection between $\alpha$-equivalence classes of
$\lambda$-terms and contextual equivalence classes of closed values of
type $\TERM$.  The Correctness of Representation property is not easy
to prove because of the nature of contextual equivalence, with its
quantification over all possible program contexts. It was established
in~\cite{ShinwellMR:freafp,PittsAM:monsf} using denotational methods
that take permutations of atoms into account. The same methods can be
used to generalise from the example of $\lambda$-terms to terms over
any \emph{nominal signature} in the sense of~\cite{PittsAM:nomu-jv}.

\subsection*{Contribution of this paper.}

For the signature in Figure~\ref{fig:signb}, the only operation on
atoms apart from $\BIND$ is a test for equality: $\atm\EQ \atm'$
evaluates to $\TRUE$ if $\atm$ and $\atm'$ are the same atom and to
$\FALSE$ otherwise. Adding extra operations and relations for atoms
may well change which program phrases are contextually equivalent.  Is
it possible to have some relations or operations on atoms in addition
to equality without invalidating the above Correctness of
Representation property? For example it would be very useful to have a
linear order $\LP\kw{<}\RP\OFTY\ATM\FUNTY\ATM\FUNTY\BOOL$, so that
values of type $\ATM$ could be used as keys in efficient data
structures for finite maps and the like. We show that this  is
possible, and more.  This is a rather unexpected result, for the
following reason.

The proof of the Correctness of Representation property given
in~\cite{ShinwellMR:freafp,PittsAM:monsf} relies upon
\emph{equivariant} properties of the semantics, in other words ones
whose truth is invariant under permuting atoms.  Atom equality is
equivariant: since a permutation is in particular bijective, it
preserves and reflects the value of $\atm\EQ \atm'$. At first it seems
that a linear order on atoms cannot be equivariant, since if
$\atm\mathbin{\kw{<}} \atm'$ is true, then applying the permutation
swapping $\atm$ and $\atm'$ we get $\atm'\mathbin{\kw{<}}\atm$, which
is false. However, equivariance is a global property: when considering
invariance of the truth of a property under permutations, it is
crucial to take into account all the parameters upon which the
property depends. Here there is a hidden parameter: \emph{the current
  state of dynamically created atoms}. So we should permute the atoms
in this state as well as the arguments of the relation. We shall see
that it is perfectly possible to have a state-dependent equivariant
ordering for the type $\ATM$ without invalidating the Correctness of
Representation property.  Indeed we prove that \emph{one can add any
  $n$-ary function from $\ATM$ to numbers \emph{(or to booleans, for
    that matter)} whose semantics is reasonable \emph{(we explain what
    is reasonable in Section~\ref{sec:observations-atoms})}, without
  invalidating the Correctness of Representation property for any
  nominal signature.}

We have to work quite hard to get this result, which generalises the
one announced in \cite{PittsAM:frepbm} (with a flawed proof sketch)
and finally proved in~\cite{PittsAM:monsf,ShinwellMR:freafp}; but
whereas those works uses denotational techniques, here we use an
arguably more direct approach based on the operational semantics of
the language. We obtain the correctness result
(Theorem~\ref{thm:corr}) as a corollary of more general result
(Propositions~\ref{prop:ext-bind-1} and \ref{prop:ext-bind-2}) showing
that, up to contextual equivalence, the type $\ty\,\BINDTY$ behaves
like the atom-abstraction construct
of~\cite[Sect.~5]{PittsAM:newaas-jv}.  Along the way to these results
we prove a Mason-Talcott-style ``CIU''~\cite{MasonIA:equfle}
characterisation of contextual equivalence for our language
(Theorem~\ref{thm:ciu}).  This is proved using Howe's
method~\cite{HoweDJ:procbf} applied to a formulation of the
operational semantics with Felleisen-style evaluation
contexts~\cite{FelleisenM:revrst}, via an abstract machine with frame
stacks~\cite{PittsAM:opespe}. The proof technique underlying our work
is rule-based induction, but with the novel twist that we exploit
semantic properties of freshness of names that are based on the use of
name permutations and that were introduced in \cite{PittsAM:newaas-jv}
and developed
in~\cite{PittsAM:nomlfo-jv,UrbanC:fortbv,PittsAM:alpsri}.

\section{Generative Unbinding}
\label{sec:generative-unbinding}

We use a version of FreshML that provides the signature in
Figure~\ref{fig:signb} in the presence of higher order recursively
defined functions on user declared data structures. Its syntax is
given in Figure~\ref{fig:lans}.

\begin{figure}\small
  \centering
  \[
  \begin{array}{lrl}
    \text{\emph{Variables}} 
    &\vid[f],\vid\in\Vid &\text{countably infinite set (fixed)}\\
    \text{\emph{Atoms}}  
    &\atm\in\Atom &\text{countably infinite set (fixed)}\\
    \text{\emph{Data types}} 
    &\dty\in\Dty &\text{finite set (variable)}\\
    \text{\emph{Constructors}} 
    &\CON\in\Con &\text{finite set (variable)}\\
    \text{\emph{Observations}} 
    &\OBS\in\Obs  &\text{finite set (variable)}\\[\jot]
    \text{\emph{Values}} 
    &\val\in\Val &\!\!\!\!::=\\
    &\makebox[0pt][r]{variable} &\vid\\
    &\makebox[0pt][r]{unit} &\UNITVAL\\
    &\makebox[0pt][r]{pair} &\PAIR{\val}{\val}\\
    &\makebox[0pt][r]{recursive function} &\FUN(\vid[f]\,\vid\EQ \exp)\\
    &\makebox[0pt][r]{data construction} &\CON\,\val\\
    &\makebox[0pt][r]{atom} &\atm\\
    &\makebox[0pt][r]{atom binding} &\BINDVAL{\val}{\val}\\[\jot]
    \text{\emph{Expressions}}
    &\exp\in\Exp &\!\!\!\!::=\\
    &\makebox[0pt][r]{value} &\val\\
    &\makebox[0pt][r]{sequencing} &\LET {\vid\EQ\exp} \IN \exp\\
    &\makebox[0pt][r]{first projection} &\FST\,\val\\
    &\makebox[0pt][r]{second projection} &\SND\,\val\\
    &\makebox[0pt][r]{function application} &\val\,\val\\
    &\makebox[0pt][r]{data deconstruction} 
    &\MATCH\  \val \WITH ({\CON\,\vid\TO\exp}\ALT\cdots)\\
    &\makebox[0pt][r]{fresh atom} &\FRESH\UNITVAL\\
    &\makebox[0pt][r]{generative unbinding} &\UNBIND\,\val\\
    &\makebox[0pt][r]{atom observation} &\OBS\,\val\cdots\val\\[\jot]
    \text{\emph{Frame stacks}} 
    &\stk\in\Stk &\!\!\!\!::=\\
    &\makebox[0pt][r]{empty} &\es\\
    &\makebox[0pt][r]{non-empty} &\stk\comp(\vid.\exp)\\
    \text{\emph{States}}
    &\s\in\State &\!\!\!\!\defeq \text{finite lists of distinct atoms}\\
    \makebox[0pt][l]{\text{\emph{Machine configurations}}}
    &&\config{\s}{\stk}{\exp}\\[\jot]
    \text{\emph{Types}} 
    &\ty\in\Ty &\!\!\!\!::=\\
    &\makebox[0pt][r]{unit} &\UNIT\\
    &\makebox[0pt][r]{pairs} &\ty\PRODTY\ty\\
    &\makebox[0pt][r]{functions} &\ty\FUNTY\ty\\
    &\makebox[0pt][r]{data type} &\dty\\
    &\makebox[0pt][r]{atoms} &\ATM\\
    &\makebox[0pt][r]{atom bindings} &\ty\,\BINDTY\\
    \makebox[0pt][l]{\text{\emph{Typing environments}}}
    &\Gamma &\!\!\!\!\!\in \Vid\fmap\Ty\\
    \makebox[0pt][l]{\text{\emph{Typing judgements}}} & &\\
    &\makebox[0pt][r]{expressions \& values}
    &\Gamma\enty\exp\ofty\ty\\
    &\makebox[0pt][r]{frame stacks}
    &\Gamma\enty\stk\ofty\ty\FUNTY\ty'\\[\jot]
    \text{\emph{Initial basis}} &&\\
    &\makebox[0pt][r]{natural numbers} &\NAT \in\Dty\\
    &\makebox[0pt][r]{zero} &(\CON[Zero]\OFTY\UNIT\FUNTY\NAT) \in\Con\\
    &\makebox[0pt][r]{successor} &(\CON[Succ]\OFTY\NAT\FUNTY\NAT) \in\Con\\
    &\makebox[0pt][r]{atom equality} & \OBS[eq] \in\Obs\quad (\arity=2)
  \end{array}
  \]
  \caption{Language syntax.}
  \label{fig:lans}
\end{figure}

\subsection*{Variable binding.}

The syntax of expressions and frame stacks in Figure~\ref{fig:lans}
involves some variable-binding constructs. Specifically:
\begin{itemize}
\item free occurrences of $\vid[f]$ and $\vid$ in $\exp$ are bound in
  $\FUN(\vid[f]\,\vid\EQ \exp)$;

\item free occurrences of $\vid$ in $\exp$ are bound in $\LET
  {\vid\EQ\exp'} \IN \exp$;

\item for $i=1..n$, free occurrences of $\vid_i$ in $\exp_i$ are bound
  in $\MATCH\ \val \WITH
  ({\CON\,\vid_1\TO\exp_1}\ALT\cdots\ALT{\CON\,\vid_n\TO\exp_n})$;

\item free occurrences of $\vid$ in $\exp$ are bound in
  $\stk\comp(\vid.\exp)$.
\end{itemize}
As usual, \emph{we identify expressions and frame stacks up to
  renaming of bound variables}. We write $\fv(\exp)$ for the finite
set of free variables of an expression $\exp$ (and similarly for frame
stacks); and we write
\begin{equation}
  \label{eq:72}
  \exp\sub{\vid,\ldots}{\val,\ldots}
\end{equation}
for the simultaneous, capture avoiding substitution of values
$\val,\ldots$ for all free occurrences of the corresponding variables
$\vid,\ldots$ in the expression $\exp$ (well-defined up to
$\alpha$-equivalence of bound variables).

\begin{figure}\small
  \centering
  \begin{align*}
    (\exp,\exp') &\defeq \LET\vid\EQ\exp\IN\LET\vid'\EQ\exp'\IN
    (\vid,\vid')
    &&(\vid\notin\fv(\exp'), \vid'\not=\vid)\\
    \lambda \vid.\,\exp &\defeq \FUN(\vid[f]\,\vid \EQ \exp)
    &&(\vid[f]\notin\fv(\exp),\vid[f]\not=\vid)\\
     k\,\exp &\defeq \LET\vid\EQ\exp\IN k\,\vid
    &&(k=\CON,\FST,\SND)\\
     \BINDVAL{\exp}{\exp'} &\defeq \LET\vid\EQ\exp\IN\LET\vid'\EQ\exp'\IN
    \BINDVAL{\vid}{\vid'}
    &&(\vid\notin\fv(\exp'), \vid'\not=\vid)\\
    \exp\,\exp' &\defeq \LET\vid\EQ\exp\IN\LET\vid'\EQ\exp'\IN
    \vid\,\vid'
    &&(\vid\notin\fv(\exp'), \vid'\not=\vid)\\
    \MATCH\ \exp \WITH (\cdots) &\defeq \LET\vid\EQ\exp\IN\MATCH\
    \vid\WITH(\cdots)
    &&(\vid\notin\fv(\cdots))\\
    \IF \exp \THEN\exp' \ELSE \exp''
    &\defeq 
    \begin{array}[t]{@{}l}
      \MATCH\ \exp \WITH {}\\
      \quad (\CON[Zero]()\TO\exp' \ALT \CON[Succ]\,\vid\TO\exp'')
    \end{array}
    &&(\vid\notin\fv(\exp'')\\
    \FRESH\ \vid\IN \exp &\defeq \LET\vid\EQ\FRESH()\IN\exp\\
    \LET {\BINDVAL{\vid_1}{\vid_2}\EQ\exp} \IN \exp' &\defeq
    \begin{array}[t]{@{}l}
      \LET\vid\EQ\exp\IN\\
      \LET\vid'\EQ\UNBIND\,\vid\IN\\
      \LET\vid_1\EQ\FST\,\vid'\IN\\
      \LET\vid_2\EQ\SND\,\vid'\IN \exp'
    \end{array}
    &&
    (\begin{array}[t]{@{}l}
      \vid,\vid'\notin\fv(\exp')\\
      \vid'\not=\vid, \vid_1\not=\vid_2)
    \end{array}\\
     \OBS\,\exp_1\cdots\exp_n &\defeq
    \begin{array}[t]{@{}l}
      \LET\vid_1\EQ\exp_1\IN\\
      \cdots\\
      \LET\vid_n\EQ\exp_n\IN\OBS\,\vid_1\cdots\vid_n
    \end{array}
    &&
    (\begin{array}[t]{@{}l}
      \vid_1,\ldots,\vid_n\notin\fv(\exp_1,\ldots,\exp_n)\\
      \vid_1,\ldots,\vid_n\ \text{distinct}).
    \end{array}
  \end{align*}
  \caption{Some ``unreduced'' forms of expression.}
  \label{fig:unrfe}
\end{figure}

\subsection*{Reduced form.}

The expressions in Figure~\ref{fig:lans} are given in a ``reduced''
form (also called ``A-normal'' form~\cite{FlanaganC:esscc}), in which
the order of evaluation is made explicit through
${\LET}$-expressions. This is not essential: the use of reduced form
makes the development of properties of the language's dynamics more
succinct and that is mostly what we are concerned with here. However,
when giving example expressions it is convenient to use the
``unreduced'' forms given in Figure~\ref{fig:unrfe}.

\begin{remark}[\textbf{Object-level binding}]
  \label{rem:objlb}
  As well as variables (standing for unknown values), the language's
  expressions and frame stacks may contain \emph{atoms} drawn from a
  fixed, countably infinite set $\Atom$. As discussed in the
  introduction, atoms are used to represent names in the object-level
  languages that are being represented as data in this programming
  meta-language. In particular a value of the form
  $\BINDVAL{\atm}{\val}$ is used to represent the object-level binding
  of a name $\atm$ in the value $\val$.  However, note that there are
  no atom-binding constructs at the programming meta-level.  The
  reader (especially one used to using lambda-abstraction to represent
  all forms of statically-scoped binding) may well ask why? Why cannot
  we factor out by $\BINDVAL{\ }{}$-bound atoms and thereby trivialise
  (one half of) the Correctness of Representation result referred to
  in the Introduction?  The reason is that it does not make semantic
  sense to try to regard $\BINDVAL{\atm}{(-)}$ as a form of meta-level
  binding and identify all expressions up to an $\alpha$-equivalence
  involving renaming $\BINDVAL{\ }{}$-bound atoms.  For example, if
  $\atm$ and $\atm'$ are two different atoms, such an
  $\alpha$-equivalence would identify
  $\FUN(\vid[f]\,\vid\EQ\BINDVAL{\atm}{\vid})$ with
  $\FUN(\vid[f]\,\vid\EQ\BINDVAL{\atm'}{\vid})$. However, these are
  two semantically different values: they are not contextually
  equivalent in the sense discussed in
  Section~\ref{sec:cont-equiv}. For example, the operational semantics
  described below gives observably different results ($0$ and $1$
  respectively) when we place the two expressions in the context
  \[
  \LET {\BINDVAL{\vid_1}{\vid_2}\EQ [-]\,\atm} \IN
  \OBS[eq]\,\vid_1\,\vid_2
  \]
  (where $\OBS[eq]\in\Obs$ is the observation for atom-equality that
  we always assume is present---see
  Remark~\ref{sec:observations-atoms}). The reason for this behaviour
  is that variables in FreshML-like languages stand for unknown values
  that may well involve atoms free at the object level. We may get
  capture of such atoms within the scope of an atom-binding
  $\BINDVAL{\atm}{(-)}$ during evaluation.  In the example, we
  replaced the hole in $[-]\,\atm$ with
  $\FUN(\vid[f]\,\vid\EQ\BINDVAL{\atm}{\vid})$ and
  $\FUN(\vid[f]\,\vid\EQ\BINDVAL{\atm'}{\vid})$ respectively, yielding
  expressions that evaluate to $\BINDVAL{\atm}{\atm}$ and
  $\BINDVAL{\atm'}{\atm}$---the first involving capture and the second
  not; and such capturing substitution does not respect naive
  $\alpha$-equivalence. So the relation of contextual equivalence that
  we define in Section~\ref{sec:cont-equiv} does not contain this
  naive $\alpha$-equivalence that identifies all (open or closed)
  expressions up to renaming of $\BINDVAL{\ }{}$-bound
  atoms.\footnote{Since the problematic possibly-capturing
    substitution is part of the dynamics of FreshML, there remains the
    possibility that the end results in the dynamics of expression
    evaluation can be made more abstract by identifying them up to
    renaming bound atoms: see Remark~\ref{rem:altos}. There are also
    less naive versions of object-level $\alpha$-equivalence that
    respect possibly-capturing substitution, such as the one developed
    in \cite{PittsAM:nomu-jv} involving hypothetical judgements about
    freshness of atoms for variables; contextual equivalence and
    ``contextual freshness'' should form a model of this notion, but
    we do not pursue this here.}  However, we will show
  (Theorem~\ref{thm:corr}) that when we restrict to closed expressions
  representing object-level languages, then contextual equivalence
  does contain (indeed, coincides with) this form of
  $\alpha$-equivalence: this is the correctness of representation
  result referred to in the Introduction.
\end{remark}

\begin{figure}\small
  \begin{mathpar}
    \inferrule{\Gamma(\vid)=\ty}{\Gamma\enty\vid\ofty\ty}
    \and
    \inferrule{\ }{\Gamma\enty\UNITVAL\ofty\UNIT}
    \and
     \inferrule{%
      \Gamma\enty\val_1\ofty\ty_1 \\ 
      \Gamma\enty\val_2\ofty\ty_2
    }{%
      \Gamma\enty\PAIR{\val_1}{\val_2}\ofty\ty_1\PRODTY\ty_2}
    \and
    \inferrule{%
      \Gamma,\vid[f]\ofty\ty\FUNTY\ty',\vid\ofty\ty\enty \exp\ofty\ty'
    }{%
      \Gamma\enty\FUN(\vid[f]\,\vid\EQ \exp)\ofty\ty\FUNTY\ty'}
    \and
    \inferrule{%
      \CON\OFTY\ty\FUNTY\dty\\
      \Gamma\enty\val\ofty\ty
    }{%
      \Gamma\enty\CON\,\val\ofty\dty}
    \and      
    \inferrule{\atm\in\Atom}{\Gamma\enty\atm\ofty\ATM}
    \and
    \inferrule{%
      \Gamma\enty\val_1\ofty\ATM \\ 
      \Gamma\enty\val_2\ofty\ty
    }{%
      \Gamma\enty\BINDVAL{\val_1}{\val_2}\ofty\ty\,\BINDTY}
    \and
    \inferrule{%
      \Gamma\enty\exp\ofty\ty\\
      \Gamma,\vid\ofty\ty\enty\exp'\ofty\ty'
    }{%
      \Gamma\enty\LET\ {\vid\EQ\exp} \IN \exp' \ofty\ty'}
    \and
    \inferrule{%
      \Gamma\enty\val\ofty\ty_1\PRODTY\ty_2
    }{%
      \Gamma\enty\FST\,\val\ofty\ty_1}
    \and
    \inferrule{%
      \Gamma\enty\val\ofty\ty_1\PRODTY\ty_2
    }{%
      \Gamma\enty\SND\,\val\ofty\ty_2}
    \and
    \inferrule{%
      \Gamma\enty\val_1\ofty\ty\FUNTY\ty'\\
      \Gamma\enty\val_2\ofty\ty
    }{%
      \Gamma\enty\val_1\,\val_2\ofty\ty'}
    \and
    \inferrule{%
      \dty \EQ \CON_1\OF\ty_1\ALT\cdots\ALT\CON_n\OF\ty_n\\
      \Gamma\enty\val\ofty\dty\\
      \Gamma,\vid_1\ofty\ty_1\enty\exp_1\ofty\ty
      \;\cdots\;
      \Gamma,\vid_n\ofty\ty_n\enty\exp_n\ofty\ty
    }{%
      \Gamma\enty\MATCH\ \val \WITH (\CON_1\,\vid_1\TO\exp_1 \ALT
      \cdots \ALT \CON_n\,\vid_n\TO\exp_n) \ofty\ty} 
    \and
    \inferrule{\ }{\Gamma\enty\FRESH\UNITVAL\ofty\ATM}
    \and
    \inferrule{%
      \Gamma\enty\val\ofty\ty\,\BINDTY
    }{%
      \Gamma\enty\UNBIND\,\val\ofty\ATM\PRODTY\ty}
    \and
    \inferrule{%
      \arity(\OBS)=k \\
      \Gamma\enty\val_1\ofty\ATM \;\cdots\; \Gamma\enty\val_k\ofty\ATM
    }{%
      \Gamma\enty\OBS\,\val_1\ldots\val_k\ofty\NAT}
    \\
    \inferrule{\ }{\Gamma\enty\es\ofty\ty\FUNTY\ty}
    \and
    \inferrule{%
      \Gamma,\vid\ofty\ty\enty \exp\ofty\ty'\\
      \Gamma\enty \stk\ofty\ty'\FUNTY\ty''
    }{%
      \Gamma\enty \stk\comp(\vid.\exp)\ofty\ty\FUNTY\ty''}
  \end{mathpar}
  Notation:
  \begin{itemize}

  \item $\Gamma,x\ofty\ty$ indicates the typing environment obtained
    by extending the finite partial function $\Gamma$ by mapping a
    variable $\vid$ to the type $\ty$ (we always assume that
    $\vid\notin\dom(\Gamma)$).
    
    \item In the typing rule for $\MATCH$-expressions, the hypothesis
    ``$\dty \EQ \CON_1\OF\ty_1\ALT\cdots\ALT\CON_n\OF\ty_n$'' refers
    to the top-level data type declaration~\eqref{eq:dec}; in other
    words, the only constructors whose result type is $\dty$ are
    $\CON_1,\ldots,\CON_n$ and $\ty_i$ is the argument type of
    $\CON_i$ (for $i=1..n$).

  \end{itemize}
  \caption{Typing relation.}
  \label{fig:typr}
\end{figure}

\subsection*{Data types and observations.}

The language defined in Figure~\ref{fig:signb} is parameterised by the
choice of a finite set $\Obs$ of function symbols that we call
\emph{observations on atoms} and whose role is discussed in
Section~\ref{sec:observations-atoms}, by a finite set $\Dty$ of
\emph{data type} symbols, and by a finite set $\Con$ of
\emph{constructor} symbols. Each constructor $\CON\in\Con$ is assumed
to come with a type, $\CON\ofty\ty\FUNTY\dty$, where $\ty\in\Ty$ and
$\dty\in\Dty$.  The choice of $\Dty$, $\Con$ and this typing
information constitutes an ML-style top-level declaration of some
(possibly mutually recursive) data types:
\begin{equation}
  \label{eq:dec}
  \begin{array}{lr@{\;}c@{\;}l}
    \TYPE & \dty_1 & \EQ & 
    \CON_{1,1}\OF\ty_{1,1} \ALT \cdots \ALT \CON_{1,n_1}\OF\ty_{1,n_1}\\
    & & \vdots & \\
    \AND & \dty_m & \EQ & 
    \CON_{m,1}\OF\ty_{m,1} \ALT \cdots \ALT \CON_{m,n_m}\OF\ty_{m,n_m}\;.
  \end{array}
\end{equation}
Here $\dty_i$ (for $i=1..m$) are the distinct elements of the set
$\Dty$ of data type symbols and $\CON_{i,j}$ (for $i=1..m$ and
$j=1..n_i$) are the distinct elements of the set $\Con$ of constructor
symbols. The above declaration just records the typing information
$\CON\OFTY\ty\FUNTY\dty$ that comes with each constructor, grouped by
result types: $\dty_i$ appears as the result type of precisely the
constructors $\CON_{i,1},\ldots,\CON_{i,n_i}$ and their argument types
are $\ty_{i,1},\ldots,\ty_{i,n_i}$. For the moment we place no
restriction on these types $\ty_{i,j}$: they can be any element of the
set $\Ty$ whose grammar is given in Figure~\ref{fig:lans}. However,
when we consider representation of object-level languages up to
$\alpha$-equivalence in Section~\ref{sec:corr-repr}, we will restrict
attention to top-level data type declarations where the types
$\ty_{i,j}$ do not involve function types.

We consider observations on atoms that return natural numbers. (The
effect of admitting some other types of operation on atoms is
discussed in Section~\ref{sec:atoa}.) So we assume $\Dty$ always
contains a distinguished data type $\NAT$ for the type of natural
numbers and that correspondingly $\Con$ contains constructors
$\CON[Zero]\ofty\UNIT\FUNTY\NAT$ and $\CON[Succ]\ofty\NAT\FUNTY\NAT$
for zero and successor.  Each $\OBS\in\Obs$ denotes a numerical
function on atoms. We assume it comes with an \emph{arity}, specifying
the number of arguments it takes: so if $\arity(\OBS)=k$ and
$(\val_1,\ldots,\val_k)$ is a $k$-tuple of values of type $\ATM$, then
$\OBS\,\val_1\ldots\val_k$ is an expression of type $\NAT$. The typing
of the language's values, expressions and frame stacks takes place in
the presence of typing environments, $\Gamma$, each assigning types to
finitely many variables.  The rules in Figure~\ref{fig:typr} for the
inductively defined typing relation are entirely standard, given that
we are following the signature in Fig~\ref{fig:signb}.

As well as an arity, we assume that each $\OBS\in\Obs$ comes with a
specified interpretation: the form this takes is discussed in 
Section~\ref{sec:observations-atoms}.

\begin{example}[\textbf{Swapping atoms}]
  \label{exa:swa}
  Examples of programming in FreshML using its characteristic feature
  of generatively unbinding atom-binding values may be found in
  \cite{PittsAM:frepbm,PittsAM:freocu}. Another feature of FreshML,
  the operation of swapping atoms, has been left out of the grammar in
  Figure~\ref{fig:lans}. However, as we mentioned in the introduction,
  there is a type-directed definition of swapping,
  $\mathit{swap}_{\ty}\OFTY\ATM\FUNTY\ATM\FUNTY\ty\FUNTY\ty$, for this
  language. For example, when $\ty$ is the type $\ATM$ of atoms we can
  make use of the observation $\OBS[eq]\in\Obs$ for atom-equality that
  we always assume is present (see
  Remark~\ref{sec:observations-atoms}) together with the
  abbreviations in Figure~\ref{fig:unrfe} and define
  \begin{equation}
    \label{eq:82}
    \mathit{swap}_{\ATM} \defeq \lambda \vid.\lambda \vid[y].
    \lambda \vid[z].\, \IF {\OBS[eq]\,\vid[z]\,\vid} \THEN \vid[y]
    \ELSE {\IF {\OBS[eq]\,\vid[z]\,\vid[y]} \THEN \vid
    \ELSE \vid[z]}.
  \end{equation}
  At unit, product, function and atom-binding types we can make use of
  standard definitions of permutation action for these types of data
  (see~\cite[Section~3]{PittsAM:alpsri}, for example):
  \begin{align}
    \label{eq:84}
    \mathit{swap}_{\UNIT} &\defeq \lambda \vid.\lambda \vid[y].
    \lambda \vid[z].\, \vid[z]\\
    \label{eq:85}
    \mathit{swap}_{\ty_1\PRODTY\ty_2} &\defeq \lambda \vid.\lambda
    \vid[y].\lambda
    \vid[z].\,(\mathit{swap}_{\ty_1}\vid\,\vid[y]\,(\FST\,\vid[z]),
    \mathit{swap}_{\ty_2}\vid\,\vid[y]\,(\SND\,\vid[z]))\\
    \label{eq:71}
    \mathit{swap}_{\ty_1\FUNTY\ty_2} &\defeq \lambda \vid.\lambda
    \vid[y].\lambda \vid[z].\lambda \vid_1.\,
    \mathit{swap}_{\ty_2}\vid\,\vid[y]\,(\vid[z]\,
    (\mathit{swap}_{\ty_1}\vid\,\vid[y]\,\vid_1))\\
    \label{eq:73}
    \mathit{swap}_{\ty\,\BINDTY} &\defeq \lambda \vid.\lambda
    \vid[y].\lambda \vid[z].\,\LET
    {\vid[z]\EQ\BINDVAL{\vid[z]_1}{\vid[z]_2}} \IN
    \BINDVAL{\mathit{swap}_{\ATM}\vid\,\vid[y]\,\vid[z]_1}
    {(\mathit{swap}_{\ty}\vid\,\vid[y]\,\vid[z]_2)}.
  \end{align}
  At data types we have to make recursive definitions corresponding to
  the inductive nature of the data types. For example, if we assume
  that in addition to the data type $\NAT$ for natural numbers we just
  have a data type $\TERM$ as in \eqref{eq:10}, then we can define
  \begin{align}
    \label{eq:74}
    \mathit{swap}_{\NAT} &\defeq \lambda \vid.\lambda \vid[y].
    \FUN(\vid[f]\,\vid[z] \EQ {\MATCH\ \vid[z]\ \WITH (\CON[Zero]()
      \TO \CON[Zero]() \ALT \CON[Succ]\,\vid[z]_1 \TO
      \CON[Succ](\vid[f]\,\vid[z]_1))})\\
    \label{eq:75}
    \mathit{swap}_{\TERM} &\defeq \lambda \vid.\lambda \vid[y].
    \FUN(\vid[f]\,\vid[z] \EQ \MATCH\ \vid[z]\ \WITH 
    \begin{array}[t]{@{}l@{}l}
      ( & \CON[V]\,\vid[z]_1 \TO 
      \CON[V](\mathit{swap}_{\ATM}\vid\,\vid[y]\,\vid[z]_1)\\
      \ALT & \CON[L]\,\vid[z]_1 \TO 
      \begin{array}[t]{@{}l}
        \LET {\BINDVAL{\vid[z]_2}{\vid[z]_3}\EQ \vid[z]_1} \IN \\
        \quad \CON[L](\BINDVAL{\mathit{swap}_{\ATM}\vid\,\vid[y]\,\vid[z]_2}
        {(f\,\vid[z]_3)})
      \end{array}\\
      \ALT & \CON[A]\,\vid[z]_1 \TO 
      \CON[A](f(\FST\,\vid[z]_1), f(\SND\,\vid[z]_1))\,)).       
      \end{array}
  \end{align}
  (The fact that values of type $\NAT$ do not involve atoms means that
  the above systematic definition of $\mathit{swap}_{\NAT}$ is in
  fact contextually equivalent to $\lambda \vid.\lambda \vid[y].
  \lambda \vid[z].\,\vid[z]$.)
\end{example}

\begin{figure}\small
  \renewcommand{\arraystretch}{1.5}
  \begin{center}
    \begin{tabular}{|c|}
      \hline
      $\config{\s}{\stk}{\exp} \trans
      \config{\s'}{\stk'}{\exp'}$\\\hline
    \end{tabular}
  \end{center}
  \begin{enumerate}
    
  \item\label{item:1} $\config{\s}{\stk\comp(\vid.\exp)}{\val}
    \trans \config{\s}{\stk}{\exp\sub{\vid}{\val}}$
    
  \item\label{item:2} $\config{\s}{\stk}{\LET{\vid\EQ\exp_1} \IN
      \exp_2} \trans \config{\s}{\stk\comp(\vid.\exp_2)}{\exp_1}$
    
  \item\label{item:3} $\config{\s}{\stk}{\MATCH\ \CON\,\val\ \WITH
      (\cdots\ALT\CON\,\vid\TO \exp\ALT\cdots)} {} \trans
    \config{\s}{\stk}{\exp\sub{\vid}{\val}}$
    
  \item\label{item:4}
    $\config{\s}{\stk}{\FST\PAIR{\val_1}{\val_2}} \trans
    \config{\s}{\stk}{\val_1}$
    
  \item\label{item:5}
    $\config{\s}{\stk}{\SND\PAIR{\val_1}{\val_2}} \trans
    \config{\s}{\stk}{\val_2}$
    
  \item\label{item:6} $\config{\s}{\stk}{\val_1\,\val_2} \trans
    \config{\s}{\stk}{\exp\sub{\vid[f],\vid}{\val_1,\val_2}}$ if
    $\val_1=\FUN(\vid[f]\,\vid\EQ \exp)$
    
  \item\label{item:7} $\config{\s}{\stk}{\FRESH\UNITVAL} \trans
    \config{\s\ords\atm'}{\stk}{\atm'}$ if
    $\atm'\notin\atoms(\s)$
    
  \item\label{item:8}
    $\config{\s}{\stk}{\UNBIND\,\BINDVAL{\atm}{\val}} \trans
    \config{\s\ords\atm'}{\stk}{\PAIR{\atm'}{\val\rename{\atm}{\atm'}}}$
    if $\atm'\notin\atoms(\s)$
    
  \item\label{item:9}\raggedright
    $\config{\s}{\stk}{\OBS\,\atm_1\ldots\atm_k} \trans
    \config{\s}{\stk}{\rep{m}}$ if $\arity(\OBS)=k$,
    $(\atm_1,\ldots,\atm_k)\in \atoms(\s)^k$ and
    $\den{\OBS}_{\s}(\atm_1,\ldots,\atm_k) = m$
    
  \end{enumerate}
  Notation:
  \begin{itemize}
    
  \item $\val\rename{\atm}{\atm'}$ is the result of replacing all
    occurrences of an atom $\atm$ by an atom $\atm'$ in the value
    $\val$;
    
  \item $\atoms(\_)$ is the finite set of all atoms occurring in
    $\_\,$;
    
  \item $\s\ords\atm'$ is the state obtained by appending an atom
    $\atm'$ not in $\atoms(\s)$ to the right of the finite list of
    distinct atoms $\s$;
    
  \item $\rep{m}$ is the the closed value of type $\NAT$ corresponding
    to $m\in\NNO$: $\rep{0}\defeq \CON[Zero]\UNITVAL$ and $\rep{m+1}
    \defeq \CON[Succ]\,\rep{m}$;
    
  \item $\den{\OBS}$ is the meaning of $\OBS$: see
    Section~\ref{sec:observations-atoms}.
    
  \end{itemize}
  \caption{Transition relation.}
  \label{fig:trar}
\end{figure}

\subsection*{Operational semantics.}

The abstract machine that we use to define the language's dynamics
has configurations of the form $\config{\s}{\stk}{\exp}$. Here $\exp$
is the expression to be evaluated, $\stk$ is a stack of evaluation
frames and $\s$ is a finite list of distinct atoms that have been
allocated so far. Figure~\ref{fig:trar} defines the transition
relation between configurations that we use to give the language's
operational semantics. The first six types of transition are all quite
standard.  Transition~\ref{item:7} defines the dynamic allocation of a
fresh atom and transition~\ref{item:8} defines generative unbinding
using a freshly created atom; we discuss transition~\ref{item:9} for
observations on atoms in the next section.  For the atom $\atm'$
in~\ref{item:7} to really be fresh, we need to know that it does not
occur in $\stk$; similarly, in~\ref{item:8} we need to know that
$\atm'$ does not occur in $(\stk,\atm,\val)$. These requirements are
met if configurations $\config{\s}{\stk}{\exp}$ satisfy that all the
atoms occurring in the frame stack $\stk$ or the expression $\exp$
occur in the list $\s$. Using the notation $\atoms(-)$ mentioned in
Figure~\ref{fig:trar}, we write this condition as
\begin{equation}
  \label{eq:64}
  \atoms(\stk,\exp)\subseteq\atoms(\s).
\end{equation}
Theorem~\ref{thm:typs} shows that this property of configurations is
invariant under transitions, as is well-typedness. Before stating this
theorem we introduce some useful terminology.

\begin{definition}[\textbf{Worlds}]
  A (\emph{possible}) \emph{world} $\w$ is just a finite subset of the
  the fixed set $\Atom$ of atoms. We write $\World$ for the set of all
  worlds.  
\end{definition}
In what follows we will index various relations associated with the
language we are considering by worlds $\w\in\World$ that make explicit
the atoms involved in the relation. Sometimes (as in the following
theorem) this is merely a matter of notational convenience;
world-indexing will be more crucial when we consider program
equivalence: see Remark~\ref{rem:posw} below.

\begin{theorem}[\textbf{Type Safety}]
  \label{thm:typs}
  Write ${}\ent_{\w}\config{\s}{\stk}{\exp}\ofty\ty$ to mean that
  $\atoms(\stk,\exp)\subseteq\atoms(\s) = \w$ and that there is some
  type $\ty'$ with $\emptyset\enty \stk\ofty\ty'\FUNTY\ty$ and
  $\emptyset\enty \exp\ofty\ty'$. The type system has the
  following properties.
  \begin{description}
  \item[Preservation] if ${}\ent_{\w}\config{\s}{\stk}{\exp}\ofty\ty$
    and $\config{\s}{\stk}{\exp}\trans \config{\s'}{\stk'}{\exp'}$,
    with $\atoms(\s')=\w'$ say, then $\w\subseteq\w'$ and
    ${}\ent_{\w'} \config{\s'}{\stk'}{\exp'}\ofty\ty$.
    
  \item[Progress] if ${}\ent_{\w}\config{\s}{\stk}{\exp}\ofty\ty$,
    then either $\stk=\es$ and $\exp\in\Val$, or
    $\config{\s}{\stk}{\exp}\trans \config{\s'}{\stk'}{\exp'}$ holds
    for some $\s'$, $\stk'$ and $\exp'$.
  \end{description}
\end{theorem}
\proof
  The proof of these properties is routine and is omitted.
\qed

\begin{remark}[\textbf{Alternative operational semantics}]
  \label{rem:altos}
  It is worth remarking that there are alternative approaches to
  representing object-level binding of a name $\atm$ in a value $\val$
  in FreshML-like languages. In the original paper on
  FreshML~\cite{PittsAM:metpbn}, the authors make a distinction
  between non-canonical expressions $\atm.\val$ for atom-binding and
  the ``semantic values'' $\mathit{abs}(\atm,val)$ to which they
  evaluate. That paper gives an operational semantics in the style of
  the Definition of Standard ML~\cite{MilnerR:defsmr} in which
  programming language expressions are separate from semantic
  values. It is possible to identify such semantic values up to
  $\alpha$-equivalence of $\mathit{abs}(\atm, {-})$-bound atoms
  without the kind of inconsistency illustrated in
  Remark~\ref{rem:objlb}. (Such semantic values in which
  $\mathit{abs}(\atm,{-})$ is a binder are used by
  Pottier~\cite{PottierF:stancf}, albeit for first-order values.)
  However, this does not help to simplify the type of Correctness of
  Representation result in which we are interested here, because
  programs are written using expressions, not semantic values. For
  example, identifying semantic values in this way,
  $\mathit{abs}(\atm,\atm)$ and $\mathit{abs}(\atm',\atm')$ are
  identical and hence trivially contextually equivalent; however the
  expressions $\atm.\atm$ and $\atm'.\atm'$ (that here we write as
  $\BINDVAL{\atm}{\atm}$ and $\BINDVAL{\atm'}{\atm'}$) are not equal
  and there is something to be done to prove that they are
  contextually equivalent. In the operational semantics
  of~\cite{PittsAM:metpbn} these expressions evaluate to the same
  semantic value up to $\alpha$-equivalence; so one would need to
  prove that contextual equivalence for that language contains
  ``Kleene equivalence''---for example by proving a ``CIU'' theorem
  like our Theorem~\ref{thm:ciu} below. So it is probably possible to
  develop the results of this paper using this slightly more abstract
  style of operational semantics with semantic values identified up to
  $\alpha$-equivalence of bound atoms. However our experience is that
  the style of operational semantics we use here, in which semantic
  values are identified with certain canonical expressions (but
  necessarily not identified up $\alpha$-equivalence of bound atoms,
  for the reasons discussed in Remark~\ref{rem:objlb}) leads to a
  simpler technical development overall.
\end{remark}


\section{Observations on Atoms}
\label{sec:observations-atoms}

\begin{figure}\small
  \renewcommand{\arraystretch}{1.5}
  \begin{center}
    \begin{tabular}{|c|c|}
      \hline
      $\config{\s}{\stk}{\exp}\terminates[n]$ &
      \strut$\config{\s}{\stk}{\exp}\terminates$\\\hline
    \end{tabular}  
  \end{center}
  \begin{mathpar}
    \inferrule{
    }{%
      \config{\s}{\es}{\val}\terminates[0]}
    \and
    \inferrule{%
      \config{\s}{\stk}{\exp}\trans\config{\s'}{\stk'}{\exp'}\\
      \config{\s'}{\stk'}{\exp'}\terminates[n]
    }{%
      \config{\s}{\stk}{\exp}\terminates[n+1]}
    \and
    \inferrule{%
      \config{\s}{\stk}{\exp}\terminates[n]
    }{%
      \config{\s}{\stk}{\exp}\terminates}
  \end{mathpar}
  \caption{Termination relations.}
  \label{fig:terr}
\end{figure}

The language we are considering is parameterised by a choice of a
finite set $\Obs$ of numerical functions on atoms.  We assume that
each $\OBS\in\Obs$ comes with a specified meaning $\den{\OBS}$. As
mentioned in the introduction, we should allow these meanings to be
dependent on the current state (the list of distinct atoms that have
been created so far). So if $\arity(\OBS)=k$, for each $\s\in\State$
we assume given a function $\den{\OBS}_{\s}:\atoms(\s)^k\fun\NNO$
mapping $k$-tuples of atoms occurring in the state $\s$ to natural
numbers.  These functions are used in the transitions of type
\ref{item:9} in Figure~\ref{fig:trar}. Not every such family
$(\den{\OBS}_{\s}\mid\s\in\State)$ of functions is acceptable as an
observation on atoms: we require that the family be
\emph{equivariant}. To explain what this means we need the following
definition.

\begin{definition}[\textbf{Permutations}]
  \label{def:perms}
  A finite \emph{permutation} of atoms is a bijection $\pi$ from the
  set $\Atom$ of atoms onto itself such that
  $\supp(\pi)\defeq\{\atm\in\Atom\mid \pi(\atm)\not=\atm\}$ is a
  finite set. We write $\Perm$ for the set of all such permutations.
  If $\pi\in\Perm$ and $\s\in\State$, then $\pi\act\s$ denotes the
  finite list of distinct atoms obtained by mapping $\pi$ over the
  list $\s$; if $\exp$ is an expression, then $\pi\act\exp$ denotes
  the expression obtained from it by applying $\pi$ to the atoms in
  $\exp$; and similarly for other syntactical structures involving
  finitely many atoms, such as values and frame stacks.
\end{definition}

We require the functions $(\den{\OBS}_{\s}\mid\s\in\State)$ associated
with each $\OBS\in\Obs$ to satisfy an \emph{equivariance} property:
for all $\pi\in\Perm$, $\s\in\State$ and
$(\atm_1,\ldots,\atm_k)\in\atoms(\s)^k$ (where $k$ is the arity of
$\OBS$)
\begin{equation}
  \label{eq:1}
  \den{\OBS}_{\s}(\atm_1,\ldots,\atm_k) = 
  \den{\OBS}_{\pi\act\s}(\pi(\atm_1),\ldots,\pi(\atm_k))\;.
\end{equation}
We impose condition~\eqref{eq:1} for the following reason.  In
Figure~\ref{fig:trar}, the side conditions on transitions of types
\ref{item:7} and \ref{item:8} do not specify which of the infinitely
many atoms in $\Atom-\atoms(\s)$ should be chosen as the fresh atom
$\atm'$. Any particular implementation of the language will make such
choices in some specific way, for example by implementing atoms as
numbers and incrementing a global counter to get the next fresh atom.
We wish to work at a level of abstraction that is independent of such
implementation details. We can do so by ensuring that we only use
properties of machine configurations $\config{\s}{\stk}{\exp}$ that
depend on the relative positions of atoms in the list $\s$, rather
than upon their identities.  So properties of configurations should be
equivariant: if $\config{\s}{\stk}{\exp}$ has the property, then so
should $\config{\pi\act\s}{\pi\act\stk}{\pi\act\exp}$ for any
$\pi\in\Perm$. The main property of configurations we need is
\emph{termination}, defined in Figure~\ref{fig:terr}, since as we see
in the next section this determines contextual equivalence of
expressions. With condition \eqref{eq:1} we have:

\begin{lemma}
  \label{lem:ter-equivar}
  If $\config{\s}{\stk}{\exp}\terminates[n]$, then
  $\config{\pi\act\s}{\pi\act\stk}{\pi\act\exp}\terminates[n]$ for any
  $\pi\in\Perm$.
\end{lemma}
\proof
  In view of the definition of termination in Figure~\ref{fig:terr}, it
  suffices to show that the transition relation is equivariant:
  \[
    {\config{\s}{\stk}{\exp}\trans
      \config{\s'}{\stk'}{\exp'}} \;\imp\;
    {\config{\pi\act\s}{\pi\act\stk}{\pi\act\exp}\trans
      \config{\pi\act\s'}{\pi\act\stk'}{\pi\act\exp'}}\;.
  \]
  This can be proved by cases from the definition of $\trans$ in
  Fig~\ref{fig:trar}.  Cases \ref{item:1}--\ref{item:8} follow from
  general properties of the action of permutations on syntactical
  structures (such as the fact that $\pi\act(e\sub{\vid}{\val})$ equals
  $(\pi\act e)\sub{\vid}{\pi\act\val}$); case \ref{item:9}
  uses property \eqref{eq:1}.
\qed

As a corollary we find that termination is indeed independent of the
choice of fresh atom in transitions of the form \ref{item:7} or
\ref{item:8}.

\begin{corollary}
  \label{cor:some-any}
  If $\config{\s}{\stk}{\FRESH}\terminates[n+1]$ with
  $\atoms(\stk)\subseteq\atoms(\s)$, then for all
  $\atm'\notin\atoms(\s)$, it is the case that
  $\config{\s\ords\atm'}{\stk}{\atm'}\terminates[n]$.  Similarly, if
  $\config{\s}{\stk}{\UNBIND\,\BINDVAL{\atm}{\val}}\terminates[n+1]$
  with $\atoms(\stk,\atm,\val)\subseteq\atoms(\s)$, then for all
  $\atm'\notin\atoms(\s)$, it is the case that
  $\config{\s\ords\atm'}{\stk}{\PAIR{\atm'}{\val\rename{\atm}{\atm'}}}\terminates[n]$.
  \qed
\end{corollary}

There are observations on atoms that are not equivariant, that is,
whose value on some atoms in a particular state does not depend just
upon the relative position of those atoms in the state. For example,
if we fix some enumeration of the set of atoms,
$\alpha:\NNO\bij\Atom$, it is easy to see that the unary observation
given by $\den{\OBS}_{\s}(\atm)=\alpha^{-1}(\atm)$ fails to satisfy
\eqref{eq:1}.  Nevertheless, there is a wide range of functions that
do have this property. Figure~\ref{fig:exaoa} gives some examples.

\begin{figure}\small
  \begin{align*}
    \intertext{\emph{Equality}, $\OBS[eq]$ ($\arity=2$):}
    \den{\OBS[eq]}_{\s}(\atm,\atm') &\;\defeq\;
    \begin{cases}
    0 &\text{if $\atm=\atm'$,}\\
    1 &\text{otherwise.}
  \end{cases}\\
  \intertext{\emph{Linear order}, $\OBS[lt]$ ($\arity=2$):}
  \den{\OBS[lt]}_{\s}(\atm,\atm') &\;\defeq\;
  \begin{cases}
    0 &\text{if $\atm$ occurs to the left of $\atm'$ in the list $\s$,}\\
    1 &\text{otherwise.}
  \end{cases}\\
  \intertext{\emph{Ordinal}, $\OBS[ord]$ ($\arity=1$):}
  \den{\OBS[ord]}_{\s}(\atm) &\;\defeq\;
  \text{$n$, if $\atm$ is the $n$th element of the list $\s$.}\\
  \intertext{\emph{State size}, $\OBS[card]$ ($\arity=0$):}
  \den{\OBS[card]}_{\s}() &\;\defeq\; \text{length of the list $\s$.}
  \end{align*}
  \caption{Examples of observations on atoms.}
  \label{fig:exaoa}
\end{figure}

\begin{remark}[\textbf{Atom-equality test}]
  \label{rem:atoet}
  The first observation on atoms given in Figure~\ref{fig:exaoa},
  $\OBS[eq]$, combined with the usual arithmetic operations for $\NAT$
  that are already definable in the language, gives us the effect of
  the function $\LP\EQ\RP\OFTY\ATM\FUNTY\ATM\FUNTY\BOOL$ from the
  signature in Figure~\ref{fig:signb}; so \emph{we assume that the set
  $\Obs$ of observations on atoms always contains $\OBS[eq]$.}
\end{remark}

\begin{remark}[\textbf{Fresh Atoms Largest}]
  \label{rem:freal}
  Note that in the operational semantics of Figure~\ref{fig:trar} we
  have chosen to make ``fresh atoms largest'', in the sense that the
  fresh atom $\atm'$ in transitions \ref{item:7} and \ref{item:8} is
  added to the right-hand end of the list $\s$ representing the current
  state. In the presence of observations on atoms other than equality,
  such a choice may well affect the properties of the notion of
  program equivalence that we explore in the next section. Other
  choices are possible, but to insist that program equivalence is
  independent of any such choice would rule out many useful
  observations on atoms (such as $\OBS[lt]$ or $\OBS[ord]$ in
  Figure~\ref{fig:exaoa}).
\end{remark}

\section{Contextual Equivalence}
\label{sec:cont-equiv}

We wish to prove that the language we have described satisfies
Correctness of Representation properties of the kind mentioned in the
introduction. To do so, we first have to be more precise about what it
means for two expressions to be \emph{contextually equivalent}, that
is, to be interchangeable in any program without affecting the
observable results of executing that program. What is a program, what
does it mean to execute it, and what results of execution do we
observe? The answers we take to these questions are: programs are
closed well-typed expressions; execution means carrying out a sequence
of transitions of the abstract machine from an initial machine
configuration consisting of a state (that is, a list of atoms
containing those mentioned in the program), the empty frame stack and
the program; and we observe whether execution reaches a terminal
configuration, that is, one of the form $\config{\s}{\es}{\val}$. We
need only observe termination because of the language's strict
evaluation strategy: observing any (reasonable) properties of the
final value $\val$ results in the same notion of contextual
equivalence. Also, it is technically convenient to be a bit more
liberal about what constitutes an initial configuration by allowing
the starting frame stack to be non-empty: this does not change the
notion of contextual equivalence because of the correspondence between
frame stacks and ``evaluation'' contexts---see the remarks after
Definition~\ref{def:cone} below.  So we can say that $\exp$ and
$\exp'$ are contextually equivalent if for all program contexts
$\mathcal{C}[-]$, the programs $\mathcal{C}[\exp]$ and
$\mathcal{C}[\exp']$ are \emph{operationally equivalent} in the
following sense.

\begin{definition}[\textbf{Operational Equivalence of Closed Expressions}]
  \label{def:opee}
  ${}\ent_{\w} \exp \opeq \exp' \ofty\ty$ is defined to hold if
  \begin{itemize}

  \item $\atoms(e,e')\subseteq\w$;

  \item $\emptyset\enty\exp\ofty\ty$ and
    $\emptyset\enty\exp'\ofty\ty$; and
    
  \item for all $\s$, $\stk$ and $\ty'$ with
    $\w\cup\atoms(\stk)\subseteq\atoms(\s)$ and
    $\emptyset\enty\stk\ofty\ty\FUNTY\ty'$, it is the case that
    $\config{\s}{\stk}{\exp}\terminates \;\bimp\;
    \config{\s}{\stk}{\exp'}\terminates$.
  \end{itemize}
\end{definition}

However, for the reasons given in \cite[Section~7.5]{PittsAM:typor},
we prefer not to phrase the formal definition of contextual
equivalence in terms of the inconveniently concrete operation of
possibly capturing substitution of open expressions for the hole
``$-$'' in program contexts $\mathcal{C}[-]$.  Instead we take the
more abstract relational approach originally advocated by
Gordon \cite{GordonAD:opeeup} and Lassen~\cite{LassenS:relrac} that
focuses upon the key features of contextual equivalence, namely that
it is \emph{the largest congruence relation for well-typed expressions
  that contains the relation of operational equivalence of
  Definition~\ref{def:opee}.} A congruence relation is an expression
relation that is an equivalence, compatible and substitutive, in the
following sense.

\begin{figure}\small
    \centering 
  \begin{mathpar}
    \inferrule{%
      \Gamma(\vid)=\ty
    }{%
      \Gamma\ent_{\w} \vid\CR{\er}\vid \ofty\ty}
    \and
    \inferrule{\ }{\Gamma\ent_{\w}\UNITVAL\CR{\er}\UNITVAL \ofty \UNIT}
    \and
    \inferrule{%
      \Gamma\ent_{\w}\val_1\er\val_1'\ofty\ty_1\\
      \Gamma\ent_{\w}\val_2\er\val_2'\ofty\ty_2
    }{%
      \Gamma\ent_{\w}\PAIR{\val_1}{\val_2} \CR{\er}
      \PAIR{\val_1'}{\val_2'} \ofty \ty_1\PRODTY\ty_2}
    \and
    \inferrule{%
      \Gamma,\vid[f]\ofty\ty\FUNTY\ty',\vid\ofty\ty\ent_{\w} 
      \exp \er \exp' \ofty\ty'
    }{%
      \Gamma\ent_{\w}\FUN(\vid[f]\,\vid\EQ \exp) \CR{\er}
      \FUN(\vid[f]\,\vid\EQ \exp') \ofty\ty\FUNTY\ty'}
    \and
    \inferrule{%
      \CON\OFTY\ty\FUNTY\dty\\
      \Gamma\ent_{\w}\val\er\val'\ofty\ty
    }{%
      \Gamma\ent_{\w}\CON\,\val\CR{\er}\CON\,\val'\ofty\dty}
    \and
    \inferrule{\atm\in\w}{\Gamma\ent_{\w}\atm\CR{\er}\atm\ofty\ATM}
    \and
    \inferrule{%
      \Gamma\ent_{\w}\val_1\er\val_1'\ofty\ATM \quad
      \Gamma\ent_{\w}\val_2\er\val_2'\ofty\ty
    }{%
      \Gamma\ent_{\w}\BINDVAL{\val_1}{\val_2} \CR{\er}
      \BINDVAL{\val_1'}{\val_2'} \ofty\ty\,\BINDTY}
    \and
    \inferrule{%
      \Gamma\ent_{\w}\exp_1\er\exp_1'\ofty\ty\\
      \Gamma,\vid\ofty\ty\ent_{\w}\exp_2\er\exp_2'\ofty\ty'
    }{%
      \Gamma\ent_{\w}\LET\ {\vid\EQ\exp_1} \IN \exp_2
      \CR{\er} \LET\ {\vid\EQ\exp_1'} \IN \exp_2' \ofty\ty'}
    \and
    \inferrule{%
      \Gamma\ent_{\w}\val\er\val'\ofty\ty_1\PRODTY\ty_2
    }{%
      \Gamma\ent_{\w}\FST\,\val\CR{\er}\FST\,\val'\ofty\ty_1}
    \and
    \inferrule{%
      \Gamma\ent_{\w}\val\er\val'\ofty\ty_1\PRODTY\ty_2
    }{%
      \Gamma\ent_{\w}\SND\,\val\CR{\er}\SND\,\val'\ofty\ty_2}
    \and
    \inferrule{%
      \Gamma\ent_{\w}\val_1\er\val_1'\ofty\ty\FUNTY\ty'\\
      \Gamma\ent_{\w}\val_2\er\val_2'\ofty\ty
    }{%
      \Gamma\ent_{\w}\val_1\,\val_2 \CR{\er} \val_1'\,\val_2' \ofty\ty'}
    \and
    \inferrule{%
      \dty \EQ \CON_1\OF\ty_1\ALT\cdots\ALT\CON_n\OF\ty_n\\
      \Gamma\ent_{\w}\val\er\val'\ofty\dty\\
      \Gamma,\vid_1\ofty\ty_1\ent_{\w}\exp_1 \er \exp_1'\ofty\ty
      \;\cdots\;
      \Gamma,\vid_n\ofty\ty_n\ent_{\w}\exp_n \er \exp_n'\ofty\ty
    }{%
      \Gamma\ent_{\w} {\MATCH\ \val \WITH (\CON_1\,\vid_1\TO\exp_1 \ALT
      \cdots \ALT \CON_n\,\vid_n\TO\exp_n)}
      \CR{\er} {\MATCH\ \val' \WITH (\CON_1\,\vid_1\TO\exp_1' \ALT
      \cdots \ALT \CON_n\,\vid_n\TO\exp_n')} \ofty\ty} 
    \and
    \inferrule{\ }{\Gamma\ent_{\w}\FRESH\UNITVAL \CR{\er} 
      \FRESH\UNITVAL\ofty\ATM}
    \and
    \inferrule{%
      \Gamma\ent_{\w}\val\er\val'\ofty\ty\,\BINDTY
    }{%
      \Gamma\ent_{\w}\UNBIND\,\val \CR{\er} 
      \UNBIND\,\val' \ofty\ATM\PRODTY\ty}
    \and
    \inferrule{%
      \arity(\OBS)=k \quad
      \Gamma\ent_{\w}\val_1 \er \val_1' \ofty\ATM 
      \;\cdots\; \Gamma\ent_{\w}\val_k \er \val_k'\ofty\ATM
    }{%
      \Gamma\ent_{\w}\OBS\,\val_1\ldots\val_k \CR{\er}
      \OBS\,\val_1'\ldots\val_k' \ofty\NAT}
       \\
    \inferrule{\ }{\Gamma\ent_{\w}\es \CR{\er} \es \ofty\ty\FUNTY\ty}
    \and
    \inferrule{%
      \Gamma,\vid\ofty\ty\ent_{\w} \exp \er \exp'\ofty\ty'\\
      \Gamma\ent_{\w} \stk \CR{\er} \stk' \ofty\ty'\FUNTY\ty''
    }{%
      \Gamma\ent_{\w} {\stk\comp(\vid.\exp)}  \CR{\er}
      {\stk'\comp(\vid.\exp')} \ofty\ty\FUNTY\ty''}  
  \end{mathpar}
  \caption{Compatible refinement $\CR{\er}$ of an expression relation $\er$.}
  \label{fig:comr}
\end{figure}

\begin{definition}[\textbf{Expression Relations}]
  \label{def:expr}
  An \emph{expression relation} $\er$ is a set of tuples
  $(\Gamma,\w,\exp,\exp',\ty)$ (made up of a typing context, a world,
  two expressions and a type) satisfying
  $\atoms(\exp,\exp')\subseteq\w$, $\Gamma\enty \exp\ofty\ty$ and
  $\Gamma\enty \exp'\ofty\ty$. We write $\Gamma\ent_{\w} \exp\er \exp'
  \ofty\ty$ to indicate that $(\Gamma,\w,\exp,\exp',\ty)$ is a member
  of $\er$.  We use the following terminology in connection with
  expression relations.
  \begin{itemize}

  \item $\er$ is an \emph{equivalence} if it is reflexive
    ($\atoms(\exp)\subseteq\w \;\conj\; \Gamma\enty\exp\ofty\ty
    \;\imp\; \Gamma\ent_{\w}\exp\er\exp\ofty\ty$), symmetric
    ($\Gamma\ent_{\w}\exp\er\exp'\ofty\ty \;\imp\;
    \Gamma\ent_{\w}\exp'\er\exp\ofty\ty$) and transitive
    ($\Gamma\ent_{\w}\exp\er\exp'\ofty\ty \;\conj\;
    \Gamma\ent_{\w}\exp'\er\exp''\ofty\ty \;\imp\;
    \Gamma\ent_{\w}\exp\er\exp''\ofty\ty$).
    
  \item $\er$ is \emph{compatible} if ${\CR{\er}}\subseteq{\er}$,
    where $\CR{\er}$ is the \emph{compatible refinement} of $\er$,
    defined in Figure~\ref{fig:comr}.
    
  \item $\er$ is \emph{substitutive} if
    $\Gamma\ent_{\w}\val\er\val'\ofty\ty \;\conj\;
    \Gamma,\vid\ofty\ty\ent_{\w}\exp\er\exp'\ofty\ty' \;\imp\;
    \Gamma\ent_{\w}
    \exp\sub{\vid}{\val}\er\exp'\sub{\vid}{\val'}\ofty\ty'$.
    
  \item $\er$ is \emph{equivariant} if
    $\Gamma\ent_{\w}\exp\er\exp'\ofty\ty \;\imp\;
    \Gamma\ent_{\pi\act\w}\pi\act\exp\er\pi\act\exp'\ofty\ty$.
    
  \item $\er$ is \emph{adequate} if
    $\emptyset\ent_{\w}\exp\er\exp'\ofty\ty\;\imp\;
    {}\ent_{\w}\exp\opeq\exp'\ofty\ty$.
  \end{itemize}
\end{definition}

We extend operational equivalence (Definition~\ref{def:opee}) to an
expression relation, $\Gamma\ent_{\w}\exp\opeqo\exp'\ofty\ty$, by
instantiating free variables with closed values:
\begin{definition}[$\opeqo$]
  \label{def:opeqo}
  Supposing $\Gamma=\{\vid_1\ofty\ty_1,\ldots,\vid_n\ofty\ty_n\}$, we
  define $\Gamma\ent_{\w}\exp\opeqo\exp'\ofty\ty$ to hold if
  \begin{itemize}

  \item $\atoms(e,e')\subseteq\w$;
    
  \item $\Gamma\enty\exp\ofty\ty$ and
    $\Gamma\enty\exp'\ofty\ty$; and
    
  \item for all $\w'\supseteq\w$ and all closed values $\val_i$ with
    $\atoms(\val_i)\subseteq\w'$ and $\emptyset\enty\val_i\ofty\ty_i$
    (for $i=1..n$), it is the case that $\ent_{\w'}
    \exp\sub{\vec{\vid}}{\vec{\val}} \opeq
    \exp'\sub{\vec{\vid}}{\vec{\val}} \ofty \ty$.
    
  \end{itemize}
  Note that for closed expressions, that is, in the case that
  $\Gamma=\emptyset$, the relation $\opeqo$ agrees with $\opeq$:
  \begin{equation}
    \label{eq:70}
    \emptyset\ent_{\w} \exp\opeqo\exp'\ofty\ty \;\bimp\;
  {}\ent_{\w}\exp\opeq\exp'\ofty\ty\;.
  \end{equation}
\end{definition}

\begin{theorem}[\textbf{CIU}]
  \label{thm:ciu}
  Operational equivalence of possibly open expressions, $\opeqo$, is a
  compatible, substitutive and adequate equivalence. It is the largest
  such expression relation. It is also equivariant.
\end{theorem}
\proof
  The fact that $\opeqo$ is equivariant follows from
  Lemma~\ref{lem:ter-equivar}. The fact that it is an equivalence and
  adequate is immediate from its definition; as is the fact that it
  contains any expression relation that is adequate, substitutive
  and reflexive. So the main difficulty is to show that it is
  compatible and substitutive. One can do this by adapting a
  construction due to Howe~\cite{HoweDJ:procbf}; see
  Appendix~\ref{app:proof-ciu-theorem}.
\qed

\begin{definition}[\textbf{Contextual Equivalence}]
  \label{def:cone}
  In view of the discussion at the beginning of this section,
  Theorem~\ref{thm:ciu} tells us that $\opeqo$ coincides with a
  conventional notion of contextual equivalence defined using program
  contexts: so from now on we refer to $\opeqo$ as \emph{contextual
    equivalence}.
\end{definition}

\begin{remark}[\textbf{Uses of closed instantiations}]
  \label{rem:ciu}
  We labelled the above theorem ``CIU'' because it is analogous to a
  theorem of that name due to Mason and Talcott~\cite{MasonIA:equfle}.
  CIU, after permutation, stands for ``Uses of Closed
  Instantiations''; and the theorem tells us that to test open
  expressions for contextual equivalence it suffices to first close
  them by substituting closed values for free variables and then test
  the resulting closed expressions for termination when they are used
  in any \emph{evaluation context}~\cite{FelleisenM:revrst}. This
  follows from the definition of $\opeqo$ and the fact that
  termination in evaluation contexts corresponds to termination of
  machine configurations via the easily verified property
  \begin{equation}
    \label{eq:2}
    \config{\s}{\stk}{\exp}\terminates \;\bimp\; 
    \config{\s}{\es}{\stk{[\exp]}}\terminates
  \end{equation}
  where the expression $\stk{[\exp]}$ is defined by recursion on the
  length of the stack $\stk$ by:
  \begin{equation}
    \label{eq:3}
    \begin{array}[c]{rcl}
      \es[\exp] &\defeq& \exp\\
      \stk\comp(\vid.\exp')[\exp] &\defeq& \stk{[\LET {\vid\EQ\exp} \IN
        \exp']}\;.
    \end{array}
  \end{equation}
\end{remark}

Theorem~\ref{thm:ciu} serves to establish some basic properties of
contextual equivalence, such as the fact that the state-independent
transitions in Figure~\ref{fig:trar} (types \ref{item:1}--\ref{item:6}
and \ref{item:9}) give rise to contextual equivalences. For example,
$\Gamma\ent_{\w} \LET {\vid\EQ\val} \IN \exp \opeqo
\exp\sub{\vid}{\val} \ofty \ty'$ holds if
$\Gamma\ent_{\w}\val\ofty\ty$ and
$\Gamma,\vid\ofty\ty\ent_{\w}\exp\ofty\ty'$. However, we have to work
a bit harder to understand the consequences of transitions of types
\ref{item:7} and \ref{item:8} for contextual equivalence at atom
binding types, $\ty\,\BINDTY$. We address this in the next section.

\begin{remark}[\textbf{Possible Worlds}]
  \label{rem:posw}
  It is immediate from the definition of $\opeqo$ that it satisfies a
  weakening property:
  \begin{equation}
    \label{eq:7}
    \Gamma\ent_{\w}\exp\opeqo\exp'\ofty\ty
    \;\conj\;
    \w\subseteq\w'
    \;\imp\;
    \Gamma\ent_{\w'}\exp\opeqo\exp'\ofty\ty\;.   
  \end{equation}
  If it also satisfied a strengthening property
  \begin{equation}
    \label{eq:6}
    \Gamma\ent_{\w'}\exp\opeqo\exp'\ofty\ty \;\conj\;
    \atoms(\exp,\exp')\subseteq\w\subseteq\w'
    \;\imp\; \Gamma\ent_{\w}\exp\opeqo\exp'\ofty\ty
  \end{equation}
  then we could make the indexing of contextual equivalence by
  ``possible worlds'' $\w$ implicit by taking $\w=\atoms(\exp,\exp')$.
  When $\Obs$ just contains $\OBS[eq]$, property \eqref{eq:6} does
  hold; this is why there is no need for indexing by possible worlds
  in~\cite{ShinwellMR:freafp,PittsAM:monsf}. However, it is not hard
  to see that the presence of some observations on atoms, such as the
  function $\OBS[card]$ in Figure~\ref{fig:exaoa}, can cause
  \eqref{eq:6} to fail. It is for this reason that we have built
  indexing by possible worlds into expression relations
  (Definition~\ref{def:expr}).
\end{remark}

\section{Correctness of Representation}
\label{sec:corr-repr}

Recall from Section~\ref{sec:generative-unbinding} that the language
we are considering is parameterised by a top-level declaration of some
(possibly mutually recursive) data types:
\begin{equation}
  \label{eq:12}
    \begin{array}{lr@{\;}c@{\;}l}
  \TYPE & \dty_1 & \EQ & 
  \CON_{1,1}\OF\ty_{1,1} \ALT \cdots \ALT \CON_{1,n_1}\OF\ty_{1,n_1}\\
  & & \vdots & \\
  \AND & \dty_m & \EQ & 
  \CON_{m,1}\OF\ty_{m,1} \ALT \cdots \ALT \CON_{m,n_m}\OF\ty_{m,n_m}\;.
  \end{array}
\end{equation}
If we restrict attention to declarations in which the argument types
$\ty_{i,j}$ of the constructors $\CON_{i,j}$ are just finite products
of the declared data types $\dty_1\ldots,\dty_m$, then the above
declaration corresponds to a \emph{many-sorted algebraic signature};
furthermore, in this case the language's values at each data type are
just the abstract syntax trees of terms of the corresponding sort in
the signature.  By allowing atoms and atom bindings in addition to
products in the argument types $\ty_{i,j}$, one arrives at the notion
of ``nominal signature'', introduced in~\cite{PittsAM:nomu-jv} and
more fully developed in~\cite{PittsAM:alpsri}. It extends the notion
of many-sorted algebraic signature with names (of possibly many kinds)
and information about name binding in constructors. Here, for
simplicity, we are restricting to a single kind of name, represented
by the type $\ATM$ of atoms; but our results extend easily to the case
of many kinds of name.

\begin{definition}[\textbf{Nominal Signatures}]
  \label{def:noms}
  The subset $\Arity\subseteq\Ty$ is given by the grammar
  \begin{equation}
    \label{eq:9}
    \ar\in\Arity 
    ::= \UNIT 
    \mid \ar\PRODTY\ar 
    \mid \dty 
    \mid \ATM 
    \mid \ar\,\BINDTY
  \end{equation}
  where $\dty$ ranges over the finite set $\Dty$ of data type symbols.
  (In other words $\Arity$ consists of those types of our language
  that do not involve any use of the function type construction,
  $\FUNTY$.)  The elements of the set $\Arity$ are called
  \emph{nominal arities}. (The notation
  $\langle\!\langle\ATM\rangle\!\rangle\ar$ is used in
  \cite{PittsAM:nomu-jv,PittsAM:alpsri} for what we here write as
  $\ar\,\BINDTY$.) A \emph{nominal signature} with a single sort of
  atoms, $\ATM$, is specified by a data type declaration~\eqref{eq:12}
  in which the argument types $\ty_{i,j}$ of the constructors
  $\CON_{i,j}$ are all nominal arities.
\end{definition}

The occurrences of $\ar\,\BINDTY$ in a nominal signature~\eqref{eq:12}
indicate arguments with bound atoms. In particular, we can associate
with each such signature a notion of \emph{$\alpha$-equivalence},
$\aeq$, that identifies closed values of nominal arity up to renaming
bound atoms. The inductive definition of $\aeq$ is given in
Figure~\ref{fig:alpe}. It generalises to an arbitrary nominal
signature the syntax-directed characterisation of $\alpha$-equivalence
of $\lambda$-terms given in \cite[p.~36]{GunterCA:sempls}. The
definition in Figure~\ref{fig:alpe} is essentially that given in
\cite{PittsAM:alpsri}, except that we have included an indexing by
possible worlds $\w$, to chime with our form of judgement for
contextual equivalence; without that indexing, the condition
``$\atm''\notin\w\supseteq\atoms(\atm,\val,\atm',\val')$'' in the rule
for $\alpha$-equivalence of values of atom binding type would be
replaced by ``$\atm''\notin\atoms(\atm,\val,\atm',\val')$''.

\begin{figure}\small
  \begin{mathpar}
    \inferrule{\ }{{}\ent_{\w}\UNITVAL\aeq\UNITVAL\ofty\UNIT}
    \and
    \inferrule{%
      {}\ent_{\w}\val_1 \aeq \val_1'\ofty\ar_1\\
      {}\ent_{\w}\val_2 \aeq \val_2'\ofty\ar_2
    }{%
      {}\ent_{\w} \PAIR{\val_1}{\val_2} \aeq 
      \PAIR{\val_1'}{\val_2'}\ofty\ar_1\PRODTY\ar_2}
    \and
    \inferrule{%
      \CON\OFTY\ar\FUNTY\dty\\
      {}\ent_{\w}\val \aeq \val' \ofty\ar
    }{%
      {}\ent_{\w} \CON\,\val \aeq \CON\,\val' \ofty\dty}
    \and
    \inferrule{%
      \atm\in\w
    }{%
      {}\ent_{\w} \atm\aeq\atm\ofty\ATM}
    \and
    \inferrule{%
      \atm''\notin\w \supseteq\atoms(\atm,\val,\atm',\val')\\
      {}\ent_{\w\cup\{\atm''\}} \val\rename{\atm}{\atm''} \aeq 
      \val'\rename{\atm'}{\atm''} \ofty \ar      
    }{%
      {}\ent_{\w}\BINDVAL{\atm}{\val} \aeq 
      \BINDVAL{\atm'}{\val'} \ofty \ar\,\BINDTY
    }
  \end{mathpar}
  \caption{$\alpha$-Equivalence.}
  \label{fig:alpe}
\end{figure}

\begin{remark}[\textbf{The role of closed values}]
  \label{rem:clov}
  For each $\ar\in\Arity$, the \emph{closed} values (that is, ones
  with no free variables) of that type, $\emptyset\ent_{\w}
  \val\ofty\ar$, correspond precisely to the ground terms (with arity
  $\ar$ and atoms in $\w$) over the given nominal signature, as
  defined in \cite{PittsAM:nomu-jv}. For example, the
  declaration~\eqref{eq:10} corresponds to the nominal signature for
  $\lambda$-calculus; and closed values of type $\TERM$ correspond as
  in \eqref{eq:11} to the abstract syntax trees for
  $\lambda$-terms---open or closed ones, with $\lambda$-calculus
  variables represented by atoms. For other examples of nominal
  signatures, with more complicated patterns of binding,
  see~\cite[Section~2.2]{PittsAM:alpsri}.

  Note that the definition of $\aeq$ in Figure~\ref{fig:alpe} cannot
  be extended naively to \emph{open} values with free variables, for
  the reasons discussed in Remark~\ref{rem:objlb}. Free variables
  stand for unknown values that may well involve atoms that get
  captured by $\BINDVAL{\ }{}$-binders upon substitution. So as we saw
  in that remark, it does not make semantic sense to say, for example,
  that $\BINDVAL{\atm}{\vid}$ and $\BINDVAL{\atm}{\vid}$ are
  $\alpha$-equivalent without putting some restrictions on the kind of
  value $\vid$ stands for. In \cite{PittsAM:nomu-jv}, Urban \emph{et
    al} consider such restrictions consisting of assumptions about the
  freshness of atoms for variables; they generalise
  Figure~\ref{fig:alpe} to a hypothetical notion of
  $\alpha$-equivalence between open values\footnote{This is a slight
    over-simplification, since their ``nominal terms'' are not just the
    open values considered here: they involved explicit
    atom-permutations as well.}, with hypotheses consisting of such
  freshness assumptions. It may be possible to relate the validity of
  this general form of $\alpha$-equivalence to contextual equivalence,
  but here we content ourselves with the following result about the
  straightforward notion of $\alpha$-equivalence on closed values
  given by Figure~\ref{fig:alpe}.
\end{remark}

\begin{theorem}[\textbf{Correctness of Representation}]
  \label{thm:corr}
  Suppose that all the observations on atoms $\OBS$ in $\Obs$ satisfy
  the equivariance property \eqref{eq:1}. For each nominal signature,
  two closed values $\val,\val'$ of the same nominal arity $\ar$ (with
  atoms contained in the finite set $\w$, say) are $\alpha$-equivalent
  if and only if they are contextually equivalent:
  \begin{equation}
    \label{eq:13}
    {}\ent_{\w} \val\aeq\val'\ofty \ar \;\bimp\; 
    {}\ent_{\w} \val \opeq\val'\ofty\ar\;.
  \end{equation}
\end{theorem}
The rest of this section is devoted to the proof of the bi-implication
in \eqref{eq:13}. Before commencing the proof we make some remarks
about the relative difficulty of each half of the bi-implication and
about alternative approaches to the proof than the one we take.

\begin{remark}[${}\ent_{\w} \val\aeq\val'\ofty \ar \;\imp\; {}\ent_{\w} \val
  \opeq\val'\ofty\ar$]
  At first sight it might seem that this implication is trivial: since
  we identify expressions up to $\alpha$-equivalence of bound
  variables, contextual equivalence automatically contains that notion
  of equivalence. However, $\aeq$ is not that meta-level
  $\alpha$-equivalence, it is $\alpha$-equivalence at the object-level
  for $\BINDVAL{\ }{}$-bound atoms. As we noted in
  Remark~\ref{rem:objlb}, identifying all (open or closed) expressions
  up to renaming $\BINDVAL{\ }{}$-bound atoms is incompatible with
  contextual equivalence: so we cannot trivialise the left-to-right
  implication in \eqref{eq:13} by factoring out in this way. Note that
  the restriction to nominal arities in Figure~\ref{fig:alpe} means
  that we do not have to consider $\aeq$ for values of the form
  $\FUN(\vid[f]\,\vid\EQ \exp)$ and hence for open expressions $\exp$
  where the naive definition of $\aeq$ would encounter the semantic
  problems discussed in Remarks~\ref{rem:objlb} and~\ref{rem:clov}.
  
  So there really is something to do to establish the left-to-right
  implication in \eqref{eq:13}. However, we will see that we have
  already done most of the heavy lifting for this half of the theorem
  by establishing the CIU Theorem~\ref{thm:ciu}.
\end{remark}

\begin{remark}[${}\ent_{\w} \val \opeq\val'\ofty\ar \;\imp\; {}\ent_{\w}
  \val\aeq\val'\ofty \ar$]
  This is equivalent to showing that if two closed values $\val$ and
  $\val'$ of nominal arity $\ar$ are not $\alpha$-equivalent, then
  they are not contextually equivalent. Proving contextual
  inequivalence is much easier than proving contextual equivalence,
  since one just has to construct a context in which the two values
  have different operational behaviour. In this case it would suffice
  to exhibit a closed expression
  $\mathit{aeq}_{\ar}\OFTY\ar\FUNTY\ar\FUNTY\NAT$ correctly
  implementing $\aeq$, in the sense that for all $\val$ and $\val'$
  
  \begin{align*}
    {}\ent_{\w} \val\aeq\val'\ofty \ar &\;\imp\; \forall
    \s.\;\w\subseteq\atoms(\s)\imp\exists\s'.\;
    \config{\s}{\es}{\mathit{aeq}_{\ar}\val\,\val'}\trans^*
    \config{\s'}{\es}{\CON[Zero]()}\\
    {}\ent_{\w} \val\not\aeq\val'\ofty \ar &\;\imp\; \forall
    \s.\;\w\subseteq\atoms(\s)\imp\exists\s'.\;
    \config{\s}{\es}{\mathit{aeq}_{\ar}\val\,\val'}\trans^*
    \config{\s'}{\es}{\CON[Succ](\CON[Zero]())}.
  \end{align*}
  It is indeed possible to construct such an expression
  $\mathit{aeq}_{\ar}$ by induction on the structure of $\ar$, by a
  definition that mimics the rules in Figure~\ref{fig:alpe}, using
  the definition of atom-swapping from Example~\ref{exa:swa} in the
  case of an atom-binding arity and using recursively defined
  functions at data types. The proof of the above properties of
  $\mathit{aeq}_{\ar}$ is relatively straightforward if tedious; one
  first has to prove suitable correctness properties for the swapping
  expressions $\mathit{swap}_{\ar}$ from Example~\ref{exa:swa}.

  This is not the route to the right-to-left implication in
  \eqref{eq:13} that we take. Instead we deduce it from a general
  ``extensionality'' property of atom-binding types $\ty\,\BIND$ that
  holds for all types $\ty$, including ones that are not nominal
  arities, that is, ones involving function types. This property
  (Propositions~\ref{prop:ext-bind-1} and \ref{prop:ext-bind-2}) shows
  that, up to contextual equivalence, the type $\ty\,\BINDTY$ behaves
  like the atom-abstraction construct
  of~\cite[Sect.~5]{PittsAM:newaas-jv}. It seems interesting in its
  own right. We are able to prove this property of general
  atom-binding types $\ty\,\BIND$ only under a restriction on
  observations on atoms over and above the equivariance
  property~\eqref{eq:1} that we always assume they possess. This is
  the ``affineness'' property given in Definition~\ref{def:affo}
  below. The equality test $\CON[eq]$ (Figure~\ref{fig:exaoa}) is
  affine and we will see that this fact is enough to prove
  Theorem~\ref{thm:corr} as stated, that is, without any restriction
  on the observations present other than equivariance.
\end{remark}

We now begin the proof of Theorem~\ref{thm:corr}.

\begin{proposition}
  \label{prop:ext}\hfill
  \begin{itemize}
    
  \item[(i)] ${}\ent_{\w}\UNITVAL\opeq\UNITVAL\ofty\UNIT$.
    
  \item[(ii)] For all types $\ty_1,\ty_2\in\Ty$, ${}\ent_{\w}
    \PAIR{\val_1}{\val_2}\opeq\PAIR{\val_1'}{\val_2'}
    \ofty\ty_1\PRODTY\ty_2$ iff ${}\ent_{\w} \val_1\opeq
    \val_1'\ofty\ty_1$ and ${}\ent_{\w} \val_2\opeq
    \val_2'\ofty\ty_2$.

  \item[(iii)] For each data type $\dty_i$ in the declaration
    \eqref{eq:12}, ${}\ent_{\w} \CON_{i,j}\,\val
    \opeq\CON_{i,j'}\,v'\ofty\dty_i$ iff $j=j'$ and ${}\ent_{\w}
    \val\opeq \val'\ofty \ty_{i,j}$.

  \item[(iv)] ${}\ent_{\w} \atm\opeq \atm'\ofty\ATM$ iff
    $\atm=\atm'\in\w$.

  \end{itemize}
\end{proposition}
\proof
  Part (i) and the ``if'' directions of (ii)--(iv) are consequences of
  the fact (Theorem~\ref{thm:ciu}) that $\opeqo$ is a compatible
  equivalence.  For the ``only if'' directions of (ii) and (iii) we
  apply suitably chosen destructors. Thus for part (ii) we use the
  operational equivalences ${}\ent_{\w}\FST\PAIR{\val_1}{\val_2} \opeq
  \val_1\ofty\ty_1$ and ${}\ent_{\w}\SND\PAIR{\val_1}{\val_2} \opeq
  \val_2\ofty\ty_2$ that are consequences of the definitions of
  $\opeq$ and the termination relation. Similarly, part (iii) follows
  from the easily established operational (in)equivalences
  \begin{align*}
    {} &\ent_{\w} \DIVERGE \not\opeq \val \ofty \ty\\
    {} &\ent_{\w} \PROJ_{i,j}\,(\CON_{i,j}\,v) \opeq v \ofty \ty_{i,j}\\
    {} &\ent_{\w} \PROJ_{i,j}\,(\CON_{i,j'}\,v) \opeq \DIVERGE \ofty
    \ty_{i,j} \quad \text{if $j\not= j'$}
  \end{align*}
  which make use of the following expressions
  \begin{align*}
    \DIVERGE &\;\defeq\; \FUN(f\,x\EQ f\,x)\UNITVAL\\
    \PROJ_{i,j}\,\val &\;\defeq\; \MATCH\ \val \WITH (\CON_{i,1}\vid_1 \TO
    \exp[d]_{j,1} \ALT \cdots \ALT \CON_{i,n_i}\,\vid_{n_i} \TO
    \exp[d]_{j,n_i})\\
    \intertext{where}
    \exp[d]_{j,j'} &\;\defeq\;
    \begin{cases}
      \vid_j &\text{if $j=j'$,}\\
      \DIVERGE &\text{if $j\not= j'$.}
    \end{cases}
  \end{align*}
  Finally, for the ``only if'' direction of part (iv) we make use of
  the fact that $\Obs$ always contains the atom equality function
  $\OBS[eq]$ from Figure~\ref{fig:exaoa}: see
  Lemma~\ref{lem:howe-4}(i) in Appendix~\ref{app:proof-ciu-theorem}.
\qed

This proposition tells us that $\opeq$ has properties mirroring those
of $\alpha$-equivalence given by the first four rules in
Figure~\ref{fig:alpe}. To complete the proof of the correctness
theorem, we need to prove a property of $\opeq$ at atom binding
arities $\ar\,\BINDTY$ that mirrors the fifth rule in that figure. We
split this into two parts, Propositions~\ref{prop:ext-bind-1} and
\ref{prop:ext-bind-2}.

\begin{proposition}
  \label{prop:ext-bind-1}
  For any type $\ty\in\Ty$, suppose we are given closed, well-typed
  atom binding values
  $\emptyset\ent_{\w}\BINDVAL{\atm}{\val}\ofty\ty\,\BINDTY$ and
  $\emptyset\ent_{\w}\BINDVAL{\atm'}{\val'}\ofty\ty\,\BINDTY$. If for
  some atom $\atm''\notin\w$ we have
  \begin{equation}
    \label{eq:14}
    {}\ent_{\w\cup\{\atm''\}}\val\rename{\atm}{\atm''} \opeq
    \val'\rename{\atm'}{\atm''} \ofty \ty
  \end{equation}
  then
  \begin{equation}
    \label{eq:15}
    {}\ent_{\w}\BINDVAL{\atm}{\val} \opeq \BINDVAL{\atm'}{\val'}
    \ofty\ty\,\BINDTY\;. 
  \end{equation}
\end{proposition}
\proof
  Unlike the previous proposition, this result is not just a simple
  consequence of the congruence properties of operational equivalence.
  It can be proved via an induction over the rules defining
  termination: see Appendix~\ref{app:proof-ext-bind-1-proposition}.
\qed

Next we need to prove the converse of the above proposition, namely
that \eqref{eq:15} implies \eqref{eq:14} for any $\atm''\notin\w$. The
difficulty is that in verifying \eqref{eq:14} we have to consider the
termination behaviour of $\val\rename{\atm}{\atm''}$ and
$\val'\rename{\atm'}{\atm''}$ in all states $\s$ with
$\atoms(\s)\supseteq\w\cup\{\atm''\}$. The atom $\atm''$ may occur at
\emph{any} position in $\s$ and not necessarily at its right-hand end;
whereas in assuming \eqref{eq:15}, all we appear to know about the
termination behaviour of $\val\rename{\atm}{\atm''}$ and
$\val'\rename{\atm'}{\atm''}$ is what happens when a fresh atom
$\atm''$ is placed at the end of the state via generative unbinding
(cf.~Remark~\ref{rem:freal}). In fact we are able to combine $\BIND$
and $\UNBIND$ operations to rearrange atoms sufficiently to prove the
result we want, but only in the presence of observations on atoms that
are insensitive to atoms being added at the left-hand (that is, least) end
of the state. The following definition makes this property of
observations precise. It uses the notation $\atm'\ords\s$ for the
state obtained from $\s\in\State$ by appending an atom $\atm'$ not in
$\atoms(\s)$ to the \emph{left} of the finite list of distinct atoms
$\s$ (cf.~$\s\ords\atm'$ defined in Figure~\ref{fig:trar}).

\begin{definition}[\textbf{Affine Observations}]
  \label{def:affo}
  An observation on atoms, $\OBS\in\Obs$, is \emph{affine} if it is
  equivariant~\eqref{eq:1} and satisfies: for all $\s\in\State$, all
  $\atm'\notin\atoms(\s)$ and all
  $(\atm_1,\ldots,\atm_k)\in\atoms(\s)^k$ (where $k$ is the arity of
  $\OBS$)
  \begin{equation}
    \label{eq:16}
    \den{\OBS}_{\atm'\ords\s}(\atm_1,\ldots,\atm_k) 
    = \den{\OBS}_{\s}(\atm_1,\ldots,\atm_k)\;. 
  \end{equation}
  For example, of the observations defined in Figure~\ref{fig:exaoa},
  $\OBS[eq]$ and $\OBS[lt]$ are affine, whereas $\OBS[ord]$ and
  $\OBS[card]$ are not.
\end{definition}

The following property of termination follows from its definition in
Figures~\ref{fig:trar} and \ref{fig:terr}, using
Corollary~\ref{cor:some-any}.

\begin{lemma}
  \label{lem:affo}
  Given a frame stack $\stk$ and an expression $\exp$, suppose that
  only affine observations on atoms occur in them. Then for all $\s$
  with $\atoms(\stk,\exp)\subseteq\atoms(\s)$ and all
  $\atm'\notin\atoms(\s)$,
  $\config{\atm\ords\s}{\stk}{\exp}\terminates[n] \;\bimp\;
  \config{\s}{\stk}{\exp}\terminates[n]$. \qed
\end{lemma}

We now give a converse of Proposition~\ref{prop:ext-bind-1}, under the
assumption that only affine observations are used. The proof is the
technically most involved result in the paper.

\begin{proposition}
  \label{prop:ext-bind-2}
  Suppose that $\Obs$ only contains affine observations.  For any type
  $\ty\in\Ty$, suppose we are given closed, well-typed atom binding
  values $\emptyset\ent_{\w}\BINDVAL{\atm}{\val}\ofty\ty\,\BINDTY$ and
  $\emptyset\ent_{\w}\BINDVAL{\atm'}{\val'}\ofty\ty\,\BINDTY$. Then
  for all atoms $\atm''\notin\w$ we have
  \begin{equation}
    \label{eq:15-1}
    {}\ent_{\w}\BINDVAL{\atm}{\val} \opeq \BINDVAL{\atm'}{\val'}
    \ofty\ty\,\BINDTY
  \end{equation}
  implies
  \begin{equation}
    \label{eq:14-1}
    {}\ent_{\w\cup\{\atm''\}}\val\rename{\atm}{\atm''} \opeq
    \val'\rename{\atm'}{\atm''} \ofty \ty\;.
  \end{equation}
\end{proposition}
\proof
  Suppose \eqref{eq:15-1} holds and that $\atm''\notin\w$. To prove
  \eqref{eq:14-1} we have to show for any $\w'\in\World$,
  $\s\in\State$ and $\ty'\in\Ty$ with
  $\atoms(\s)=\w'\supseteq\w\cup\{a''\}$ and
  $\emptyset\ent_{\w'}\stk\ofty\ty\FUNTY\ty'$ that
  \begin{equation}  
    \label{eq:17}
    \config{\s}{\stk}{\val\rename{\atm}{\atm''}}\terminates \;\bimp\;
    \config{\s}{\stk}{\val'\rename{\atm'}{\atm''}}\terminates\;.
  \end{equation}
  Since $a''\in \atoms(\s)$, we have
  \begin{equation}
    \label{eq:18}
    \s = \s'\ords \atm''\ords \atm_0\ords\cdots\ords \atm_{n-1}
  \end{equation}
  for some state $\s'$ and atoms $\atm_0,\ldots,\atm_{n-1}$ ($n\geq
  0$).  Choose distinct atoms $\atm[b]_0,\ldots,\atm[b]_{n-1}$ not
  occurring in $\w'$ and consider the frame stack
  \begin{equation}
    \label{eq:20}
    \begin{array}{l@{}l}
      \stk' \;\defeq\; 
      \es\comp (\vid[z].\, &\LET {\BINDVAL{\vid}{\vid[y]_0}\EQ \vid[z]} \IN\\
      &\LET {\BINDVAL{\vid_0}{\vid[y]_1}\EQ\BINDVAL{\atm[b]_0}{\vid[y]_0}} \IN\\
      &\;\vdots\\
      &\LET {\BINDVAL{\vid_{n-1}}{\vid[y]_n}\EQ
        \BINDVAL{\atm[b]_{n-1}}{\vid[y]_{n-1}}} \IN{}\\
      &\stk\rename{\atm'',\atm_0\ldots,\atm_{n-1}}%
      {\vid,\vid_0,\ldots,\vid_{n-1}}[\vid[y]_n])
    \end{array}
  \end{equation}
  where
  $\vid[z],\vid,\vid_0,\ldots,\vid_{n-1},\vid[y]_0,\ldots,\vid[y]_n$
  are distinct variables not occurring in $\stk$. Here we have used
  the notation ``$\LET {\BINDVAL{\vid_1}{\vid_2}\EQ\exp} \IN \exp' $''
  from Figure~\ref{fig:unrfe}, the notation ``$\stk{[\exp]}$'' from
  \eqref{eq:3} and the operation $(-)\rename{\atm}{\vid}$ of replacing
  an atom $\atm$ by a variable $\vid$.

  Since $\atoms(\stk)\subseteq\w'=\atoms(\s)$, by definition of
  $\stk'$ and from \eqref{eq:18} we have
  $\atoms(\stk')\subseteq\atoms(\s[b]')$ where
  \begin{equation}
    \label{eq:69}
    \s[b]' \;\defeq\; \atm[b]_0\ords\cdots\ords\atm[b]_{n-1}\ords\s'\;.
  \end{equation}
  Let $\pi\in\Perm$ be the permutation swapping each $\atm_i$ with
  $\atm[b]_i$ (for $i=0..n-1$). Since
  $\atm''\notin\w\supseteq\atoms(\atm,\val)$, by definition of
  $\s[b]'$ we have
  $\atoms(\pi\act\BINDVAL{\atm}{\val})\subseteq\atoms(\s[b]')$.
  Therefore the configuration
  $\config{\s[b]'}{\stk'}{\pi\act\BINDVAL{\atm}{\val}}$ satisfies the
  well-formedness condition needed to apply
  Corollary~\ref{cor:some-any}. Noting that $\pi\act
  (\BINDVAL{\atm}{\val}) = \BINDVAL{\pi(\atm)}{(\pi\act\val)}$ and
  that $\pi\act(\val\rename{\atm}{\atm''}) =
  (\pi\act\val)\rename{\pi(\atm)}{\pi(\atm'')} = (\pi\act
  \val)\rename{\pi(\atm)}{\atm''}$, from that corollary, property
  \eqref{eq:2} and the definition of $\stk'$ we get:
  \begin{multline*}
  \config{\s[b]'}{\stk'}{\pi\act(\BINDVAL{\atm}{\val})}\terminates
    \;\bimp\;\\
    \config{\s[b]'\ords\atm''\ords\atm_0\ords\cdots\ords\atm_{n-1}}{\stk}
    {(\pi\act(\val\rename{\atm}{\atm''}))
      \rename{\atm[b]_0,\ldots,\atm[b]_{n-1}}
      {\atm_0,\ldots,\atm_{n-1}}}\terminates\;.
  \end{multline*}
  Note that by definition of $\pi$
  \begin{align*}
          &(\pi\act(\val\rename{\atm}{\atm''}))
          \rename{\atm[b]_0,\ldots,\atm[b]_{n-1}}
          {\atm_0,\ldots,\atm_{n-1}}\\
    {}={} &((\val\rename{\atm}{\atm''})
    \rename{\atm_0,\ldots,\atm_{n-1}}
    {\atm[b]_0,\ldots,\atm[b]_{n-1}})
    \rename{\atm[b]_0,\ldots,\atm[b]_{n-1}}
    {\atm_0,\ldots,\atm_{n-1}}\\
    {}={} &\val\rename{\atm}{\atm''}\;;
  \end{align*}
  and $\s[b]'\ords \atm''\ords \atm_0\ords\cdots\ords \atm_{n-1} =
  \atm[b]_0\ords\cdots\ords\atm[b]_{n-1}\ords\s$ by \eqref{eq:18} and
  \eqref{eq:69}. So altogether we have
  \begin{equation}
    \label{eq:23}
    \config{\s[b]'}{\stk'}{\pi\act\BINDVAL{\atm}{\val}}\terminates
    \;\bimp\;
    \config{\atm[b]_0\ords\cdots\ords\atm[b]_{n-1}\ords\s}%
    {\stk}{\val\rename{\atm}{\atm''}}\terminates\;.  
  \end{equation}
  A similar argument gives
  \begin{equation}
    \label{eq:24}
    \config{\s[b]'}{\stk'}{\pi\act\BINDVAL{\atm'}{\val'}}\terminates
    \;\bimp\;
    \config{\atm[b]_0\ords\cdots\ords\atm[b]_{n-1}\ords\s}%
    {\stk}{\val'\rename{\atm'}{\atm''}}\terminates\;.  
  \end{equation}
  We noted in Theorem~\ref{thm:ciu} that operational
  equivalence is equivariant. So from \eqref{eq:15-1} we have
  ${}\ent_{\atoms(\s[b]')}\pi\act\BINDVAL{\atm}{\val} \opeq
  \pi\act\BINDVAL{\atm'}{\val'} \ofty \ty\BINDTY$. Since
  $\emptyset\ent_{\atoms(\s[b]')} \stk'\ofty\ty\,\BINDTY\FUNTY\ty'$,
  this operational equivalence gives
  \[
  \config{\s[b]'}{\stk'}{\pi\act\BINDVAL{\atm}{\val}}\terminates
  \;\bimp\;
  \config{\s[b]'}{\stk'}{\pi\act\BINDVAL{\atm'}{\val'}}\terminates\;.
  \]
  Combining this with \eqref{eq:23} and \eqref{eq:24} yields
  \begin{equation}
    \label{eq:22}
    \config{\atm[b]_0\ords\cdots\ords\atm[b]_{n-1}\ords\s}%
    {\stk}{\val\rename{\atm}{\atm''}}\terminates
    \;\bimp\;
    \config{\atm[b]_0\ords\cdots\ords\atm[b]_{n-1}\ords\s}%
    {\stk}{\val'\rename{\atm'}{\atm''}}\terminates\;.
  \end{equation}
  Since $\atm[b]_0,\ldots,\atm[b]_{n-1}\notin \w' =
  \atoms(\s)\supseteq\atoms(\stk,\atm'',\val,\val')$ and $\Obs$ only
  contains affine observations, we can now apply Lemma~\ref{lem:affo}
  to \eqref{eq:22} to get \eqref{eq:17}, as required.
\qed

\begin{example}
  \label{exa:conjecture}
  We conjecture that Proposition~\ref{prop:ext-bind-2} fails to hold
  if we drop the requirement that observations are affine (but still
  require them to be equivariant). For example consider the
  equivariant but non-affine observation $\OBS[ord]$ in
  Figure~\ref{fig:exaoa} and the values
  \begin{align*}
    \val &\;\defeq\; \FUN(\vid[f]\,\vid\EQ\vid[f]\,\vid)\\
    \val' &\;\defeq\; \FUN(\vid[f]\,\vid\EQ \MATCH\ {\OBS[ord]\,\atm}
    \WITH (\CON[Zero]\TO\UNITVAL \ALT
    \CON[Succ]\,\vid[y]\TO\val\UNITVAL))
  \end{align*}
  where $\atm$ is some atom. We claim that
  \begin{align}
    {} &\ent_{\{\atm\}} \BINDVAL{\atm}{\val} \opeq
    \BINDVAL{\atm}{\val'}
    \ofty (\UNIT\FUNTY\UNIT)\BINDTY\label{eq:19}\\
    \intertext{but that for any $\atm'\not=\atm$} {}
    &\ent_{\{\atm,\atm'\}} \val\rename{\atm}{\atm'}\not\opeq
    \val'\rename{\atm}{\atm'}\ofty \UNIT\FUNTY\UNIT\;.\label{eq:21}
  \end{align}
  The operational inequivalence \eqref{eq:21} is witnessed by the
  state $\s \defeq [a',a]$ and the frame stack $\stk\defeq
  \es\comp(\vid.\,\vid\,\UNIT)$, for which one has
  $\config{\s}{\stk}{\val'\rename{\atm}{\atm'}}\terminates$, but not
  $\config{\s}{\stk}{\val\rename{\atm}{\atm'}}\terminates$. At the
  moment we lack a formal proof of the operational equivalence
  \eqref{eq:19}, but the intuitive justification for it is as follows.
  For any state $\s$ containing $\atm$ and any frame stack $\stk$, we
  claim that in any sequence of transitions from
  $\config{\s}{\stk}{\BINDVAL{\atm}{\val'}}$ the occurrence of
  $\OBS[ord]\,\atm$ in $\val'$ can only be renamed to
  $\OBS[ord]\,\atm'$ for atoms $\atm'$ at positions strictly greater
  than $0$ in the current state; and hence
  $\config{\s}{\stk}{\BINDVAL{\atm}{\val'}}$ has the same termination
  properties as $\config{\s}{\stk}{\BINDVAL{\atm}{\val}}$.
\end{example}

\proof[Proof of Theorem~\ref{thm:corr}]
  One proves that ${}\ent_{\w} \val\aeq\val'\ofty \ar$ implies
  ${}\ent_{\w} \val \opeq\val'\ofty\ar$ by induction on the the rules
  defining $\alpha$-equivalence in Figure~\ref{fig:alpe}, using
  Propositions~\ref{prop:ext} and \ref{prop:ext-bind-1}.

  To prove the converse implication, first note that if $\emptyset\ent
  \val\ofty\ar$, then $\val$ contains no instances of observations
  $\OBS\in\Obs$. The proof of this is by induction on the structure of
  the nominal arity $\ar$; the only way observations on atoms can
  appear in values of the language is via function values,
  $\FUN(\vid[f]\,\vid\EQ\exp)$, and the definition of ``nominal
  arity'' excludes function types. It follows from the definition of
  operational equivalence in Definition~\ref{def:opee} that if
  ${}\ent_{\w} \val \opeq\val'\ofty\ar$ holds for a language with
  observation set $\Obs$, it also holds for the sub-language with
  minimal observation set $\{\OBS[eq]\}$. Thus it suffices to prove
  the implication ${}\ent_{\w} \val \opeq\val'\ofty\ar \;\imp\;
  {}\ent_{\w} \val\aeq\val'\ofty \ar$ for this minimal sub-language;
  and this can be done by induction on the structure of $\ar$ using
  Propositions~\ref{prop:ext} and \ref{prop:ext-bind-2} (the latter
  applies because $\OBS[eq]$ is affine).
\qed

\section{Related and Further Work}
\label{sec:related-work}

\subsection{Correctness of Representation}

It is instructive to compare the Correctness of Representation
property of FreshML (Theorem~\ref{thm:corr}) with \emph{adequacy}
results for type-theoretic logical frameworks~\cite{PfenningF:logf}.
Both are concerned with the representation of expressions of some
object-language in a meta-language. For logical frameworks the main
issue is surjectivity: one wants every expression at the meta-level to
be convertible to a normal form and for every normal form at certain
types to be the representation of some object-level expression. The
fact that $\alpha$-equivalence of object-level expressions is
preserved and reflected by the representation is a simple matter,
because equivalence in the logical framework is taken to be
$\alpha\beta\eta$-conversion, which specialises on normal forms to
just $\alpha$-equivalence. Contrast this with the situation for
FreshML where surjectivity of the representation is straightforward,
because values of the relevant FreshML data types \emph{are} just
first order abstract syntax trees; whereas the fact that
$\alpha$-equivalence of object-level expressions is preserved and
reflected by the representation in FreshML is a non-trivial property.
This is because we take equivalence of FreshML expressions to be
contextual equivalence. This is the natural notion of equivalence from
a programming point of view, but its properties are hard won.

One aspect of adequacy results for logical frameworks highlighted in
\cite{PfenningF:logf} is \emph{compositionality} of representations.
Although important, this issue is somewhat orthogonal to our concerns
here. It refers to the question of whether substitution of expressions
for variables at the object-level is represented by $\beta$-conversion
at the meta-level. From the point of view of nominal
signatures~\cite{PittsAM:alpsri}, variables are just one kind of name.
Properties of $\alpha$-conversion of all kinds of names are treated by
the theory; but if one wants notions of substitution and
$\beta$-conversion for a particular kind of name, one has to give a
definition (an ``$\alpha$-structural'' recursive
definition~\cite{PittsAM:alpsri}). For example in FreshML, using the
data type \eqref{eq:10} for $\lambda$-terms one can give an
appealingly simple declaration for a function $\kw{subst}\OFTY
\TERM\FUNTY\ATM\FUNTY\TERM\FUNTY\TERM$ for capture-avoiding
substitution; see~\cite[p.~264]{PittsAM:frepbm}.  Compositionality of
the representation $t\mapsto \rep{t}$ given in the introduction then
becomes the contextual equivalence ${}\ent_{\w}
\rep{t_1\sub{\atm}{t_2}} \opeq \kw{subst}\,\rep{t_2}\,\atm\,\rep{t_1}
\ofty\TERM$. The CIU theorem (Theorem~\ref{thm:ciu}) provides the
basis for proving such contextual equivalences. (We believe this
particular equivalence is valid when $\Obs=\{\OBS[eq],\OBS[lt]\}$, but
not when $\Obs=\{\OBS[eq],\OBS[card]\}$; see
Section~\ref{sec:conclusion}.)

\subsection{Concrete Semantics}
\label{sec:atoa}

We have explored some of the consequences of adding integer-valued
``observations on atoms'' to FreshML over and above the usual test
for equality.  Such functions allow more efficient data structures
to be used for algorithms involving atoms as keys. For example, binary
search trees making use of the comparison function $\OBS[lt]$ from
Figure~\ref{fig:exaoa} could be used instead of association lists.

What about adding functions from numbers to atoms?  An implementation
of the language may well represent atoms by numbers, via some fixed
enumeration of the set of atoms, $\alpha:\NNO\bij\Atom$.  Can we give
the programmer access to this bijection? Less radically, can we allow
operations on atoms that make use of arithmetic properties of the
underlying representation? Not without breaking the invariant
$\atoms(\stk,\exp)\subseteq\atoms(\s)$ of configurations
$\config{\s}{\stk}{\exp}$---the property of our operational semantics
that ensures that an atom's freshness with respect to the current
state really does mean that it is different from all other atoms in
the current context.  For example, suppose we add to the language an
operation $\OBS[suc]\OFTY\ATM\FUNTY\ATM$ whose meaning is ``successor
function on atoms'', with transitions
$\config{\s}{\stk}{\OBS[suc]\,\atm} \trans \config{\s}{\stk}{\atm'}$
whenever $\atm=\alpha(n)$ and $\atm'=\alpha(n+1)$ for some $n\in\NNO$.
Then it may well be the case that $\atm'\notin\atoms(\s)$ even though
$\atm\in\atoms(\s)$.

So exposing the numerical representation of atoms involves giving up
the invariant properties of the abstract semantics we have used here.
Perhaps a more interesting alternative to actually
exposing numerical representations of atoms would be to prove
contextual equivalence of efficient and naive implementations of the
abstract semantics extended with types of finite maps on atoms.  Such
abstract types form an addition to the signature in
Figure~\ref{fig:signb} different from the kind we have considered
here, but certainly one worthy of investigation.

\subsection{Mechanising Meta-Theory}

The techniques we used here to prove the Correctness of Representation
property are operationally based, in contrast to the denotational
techniques used in~\cite{ShinwellMR:freafp,PittsAM:monsf}. The
advantage of working directly with the syntax and operational
semantics of the language is that there are lower mathematical
``overheads''---various kinds of induction being the main techniques
involved. The disadvantage is that to deploy such inductive techniques
often involves great ingenuity choosing inductive hypotheses and much
error prone tedium checking induction steps. Furthermore, with these
methods it seems harder to predict the effect that a slight change in
language or formalisation may have on a proof. Although ingenuity in
choosing inductive hypotheses may always be the preserve of humans,
machine assistance of the kind envisaged by the ``POPLmark
challenge''~\cite{PierceBP:mecmmp} seems a very good idea for the
other aspects of the operationally based approach.  The main results
presented here are still a challenging target for fully formalised and
machine checked proofs. We have taken some care with the formalisation
(using a ``relational'' approach to contextual equivalence, for
example); but results concerning coinductive equivalences, like the
CIU theorem (Theorem~\ref{thm:ciu}), are quite complex logically
speaking, compared with the kind of type safety results (like
Theorem~\ref{thm:typs}) that POPLMark has focused on so far.  Systems
like Isabelle/HOL~\cite{NipkowT:isahpa} that develop proofs in full
classical higher order logic seem appropriate to the task, in
principle. But there is a gap between what is possible in principle
for an expert of any particular system and what is currently
practicable for a casual user. Urban and
Berghofer~\cite{UrbanC:reccnd} are developing a \emph{Nominal Data
  Type Package} for Isabelle/HOL that may be very useful for narrowing
this gap. The fact that FreshML and the Urban-Berghofer package both
have to do with the same mathematical universe of ``nominal
sets''~\cite{PittsAM:alpsri} is perhaps slightly confusing: their
Nominal Data Type Package is useful for fully formalising proofs about
names and name-binding in operational semantics whether or not those
proofs have to do with the particular mechanism of generative
unbinding that is the focus of this paper.

\section{Conclusion}
\label{sec:conclusion}

The FreshML~\cite{PittsAM:frepbm,ShinwellMR:freona} approach to
functional programming with binders combines abstract types for names
and name binding with an unbinding operation that involves generation
of fresh names.  In this paper we have studied some theoretical
properties of this design to do with data correctness. We showed that
the addition of integer valued observations on names does not break
FreshML's fundamental Correctness of Representation property that
$\alpha$-equivalence classes of abstract syntax trees (for any nominal
signature) coincide with contextual equivalence classes of user
declared data values. In particular, it is possible to give
programmers access to a linear order on names without breaking the
``up to $\alpha$-equivalence'' representation of syntax. The simple
insight behind this possibly surprising result has to do with the fact
that FreshML is impure---program execution mutates the state of
dynamically created names. If the state is taken into account when
giving the meaning of observations on names, then the permutation
invariance properties that underly the correctness property can be
retained.  The original version of FreshML~\cite{PittsAM:metpbn} was
pure by dint of the ``freshness inference'' included in its type
system. Subsequent experience with the language showed that the form
of freshness inference that was used there was overly restrictive from
a programming point of view. So freshness inference was dropped in
\cite{PittsAM:frepbm}.  However, Pottier~\cite{PottierF:stancf} has
recently regained purity in a FreshML-like language through the use of
user-provided assertions. We have not investigated whether results
like those presented in this paper also apply to Pottier's language.

This paper has been concerned with data correctness, but what about
general results on \emph{program correctness}? The only restriction we
placed on observations on atoms is that, as functions of both the
state and the names they operate upon, they should be invariant under
permuting names. We have seen that the Correctness of Representation
property (Theorem~\ref{thm:corr}) remains valid in the presence of any
such observation. However, we are certainly not advocating that
arbitrary equivariant observations be added to FreshML. This is
because some forms of observation may radically affect the general
programming laws that contextual equivalence satisfies.  We saw one
example of this here: only for ``affine'' observations (which are
insensitive to how many names have been created before the arguments
to which they are applied) were we able to combine
Propositions~\ref{prop:ext-bind-1} and~\ref{prop:ext-bind-2} to get an
``extensionality'' result explaining contextual equivalence at type
$\ty\,\BINDTY$ in terms of contextual equivalence at $\ty$, for
arbitrary higher types $\ty$.

More investigation of program correctness properties in the presence
of particular observations on atoms is needed before one can advocate
adding them to the FreshML design.  The techniques we used in this
paper could form the basis for such an investigation. They combine the
usual engine of structural operational semantics---namely
syntax-directed, rule based induction---with the approach to freshness
of names based on name permutations that was introduced in
\cite{PittsAM:newaas-jv} and developed
in~\cite{PittsAM:nomlfo-jv,UrbanC:fortbv,PittsAM:alpsri}.

\section*{Acknowledgement.}

The authors are grateful for the suggestions for improvement made by
the anonymous referees.

\newcommand{\etalchar}[1]{$^{#1}$}

\appendix
\section{Proof of Theorem~\ref{thm:ciu}}
\label{app:proof-ciu-theorem}

We wish to show that the expression relation $\opeqo$ of
Definition~\ref{def:opeqo} is compatible and substitutive (see
Definition~\ref{def:expr}). We use an adaptation of ``Howe's
method''~\cite{HoweDJ:procbf} to do so. Let the expression relation
$\opeq^*$ be inductively defined from $\opeqo$ by the rule
\begin{equation}
  \label{eq:4}
  \inferrule{%
    \Gamma\ent_{\w} \exp \CR{\opeq^*} \exp' \ofty\ty\\
    \Gamma\ent_{\w} \exp' \opeqo \exp'' \ofty \ty
  }{%
    \Gamma\ent_{\w} \exp \opeq^* \exp''\ofty\ty
  }\;.
\end{equation}
In making this inductive definition, we are implicitly relying upon
the easily proved fact that compatible refinement, ${\er}\mapsto
{\CR{\er}}$, is a monotone operation on expression relations,
that is, ${\er_1}\subseteq{\er_2} \;\imp\;
{\CR{\er_1}}\subseteq{\CR{\er_2}}$.

\begin{lemma}
  \label{lem:howe-1}\hfill
  \begin{itemize}
  \item[(i)] $\Gamma\ent_{\w}\exp\opeq^*\exp'\ofty\ty \;\conj\;
    \Gamma\ent_{\w}\exp'\opeqo\exp''\ofty\ty \;\imp\;
    \Gamma\ent_{\w}\exp\opeq^*\exp''\ofty\ty$.

  \item[(ii)] $\opeq^*$ is compatible and substitutive.

  \item[(iii)] $\atoms(\exp)\subseteq\w \;\conj\;
    \Gamma\enty\exp\ofty\ty \;\imp\;
    \Gamma\ent_{\w}\exp\opeq^*\exp\ofty\ty$.

  \item[(iv)] $\atoms(\stk)\subseteq\w \;\conj\;
    \Gamma\enty\stk\ofty\ty\FUNTY\ty' \;\imp\; \Gamma\ent_{\w}\stk
    \CR{\opeq^*}\stk \ofty\ty\FUNTY\ty'$.

  \item[(v)] $\Gamma\ent_{\w}\val\opeq^*\exp'\ofty\ty \;\imp\;\exists
    \val'.\;\Gamma\ent_{\w}\val\opeq^* \val'\ofty\ty \;\conj\;
    \Gamma\ent_{\w}\val'\opeqo\exp'\ofty\ty$.
  \end{itemize}
\end{lemma}
\proof
  These properties of $\opeq^*$ are simple consequences of its
  definition and the definition of the extension of compatible
  refinement to a relation between frame stacks given by the last two
  rules in Figure~\ref{fig:comr}.
\qed

\begin{lemma}
  \label{lem:howe-2}\hfill
  \begin{itemize}
  \item[(i)] $\opeq^*$ is equivariant.

  \item[(ii)] $\Gamma\ent_{\w}\exp\opeq^*\exp'\ofty\ty \;\conj\;
    \w\subseteq\w' \;\imp\; \Gamma\ent_{\w'}\exp\opeq^*\exp'\ofty\ty$.

  \item[(iii)] $\Gamma\ent_{\w}\stk \CR{\opeq^*} \stk'
    \ofty\ty\FUNTY\ty' \;\conj\; \w\subseteq\w' \;\imp\;
    \Gamma\ent_{\w'}\stk\CR{\opeq^*} \stk' \ofty\ty\FUNTY\ty'$.
  \end{itemize}
  
\end{lemma}
\proof
  Part (i) follows from the fact that $\opeqo$ is equivariant, which
  in turn is a consequence of Lemma~\ref{lem:ter-equivar}. Parts (ii)
  and (iii) are consequences of the fact that world weakening is built
  into the definition of operational equivalence in
  Definition~\ref{def:opee}.
\qed

\begin{lemma}
  \label{lem:howe-3}
  $\Gamma\ent_{\w}\exp\opeqo\exp'\ofty\ty \;\imp\;
  \Gamma\ent_{\w}\exp\opeq^*\exp'\ofty\ty$.
\end{lemma}
\proof
  If $\Gamma\ent_{\w}\exp\opeqo\exp'\ofty\ty$, then in particular
  $\atoms(\exp)\subseteq\w$ and $\Gamma\enty \exp\ofty\ty$, so that by
  Lemma~\ref{lem:howe-1}(iii) we have $\Gamma\ent_{\w} \exp \opeq^*
  \exp\ofty\ty$; so from part (i) of that lemma we get
  $\Gamma\ent_{\w} \exp \opeq^* \exp'\ofty\ty$.
\qed

We wish to show that $\opeq^*$ coincides with $\opeqo$. In view of the
previous lemma, it just remains to show that
${\opeq^*}\subseteq{\opeqo}$.  Lemma~\ref{lem:howe-key} provides the
key to this. Before stating that lemma we give some simple properties
of $\opeq$ that are needed to prove it.

\begin{lemma}
  \label{lem:howe-4}\hfill
  \begin{itemize}
  \item[(i)] ${}\ent_{\w}\atm\opeq\atm'\ofty\ATM \;\imp\; \atm=\atm'$.

  \item[(ii)] ${}\ent_{\w}\val\opeq\val'\ofty\ty\,\BINDTY \;\imp\;
    {}\ent_{\w}\UNBIND\,\val\opeq\UNBIND\,\val'\ofty\ATM\PRODTY\ty$.

  \item[(iii)] If ${}\ent_{\w}\val\opeq\val'\ofty\ty_1\FUNTY\ty_2$,
    then for any world $\w'\supseteq\w$ and value $\val_1$ with
    $\atoms(\val_1)\subseteq\w'$ and ${}\enty\val_1\ofty\ty_1$, it is
    the case that
    ${}\ent_{\w'}\val\,\val_1\opeq\val'\,\val_1\ofty\ty_2$.
  \end{itemize}
\end{lemma}
\proof
  For part (i) we make use of the fact that $\Obs$ always contains the
  atom equality function $\OBS[eq]$ from Figure~\ref{fig:exaoa}.
  Consider the frame stack
  \[
  \stk_{\atm} \;\defeq\; \es\comp(\vid.\,
  \begin{array}[t]{@{}l@{}}
    \LET {\vid[y]\EQ\OBS[eq]\,\vid\,\atm} \IN\\
    \MATCH\ \vid[y] \WITH 
    (\CON[Zero]\TO \UNITVAL\ALT \CON[Succ]\,\vid[z]\TO\DIVERGE))\;.
  \end{array}
  \]
  If $\atm\not=\atm'$ are distinct elements of $\w$, then choosing
  some $\s\in\State$ with $\atoms(\s)=\w$, it is not hard to see that
  $\config{\s}{\stk_{\atm}}{\atm}\terminates$ holds whereas
  $\config{\s}{\stk_{\atm}}{\atm'}\terminates$ does not hold. So if
  ${}\ent_{\w} \atm\opeq \atm'\ofty\ATM$ it cannot be the case that
  $\atm\not= \atm'$.
    
  For part (ii), given any $\s$, $\stk$ and $\ty'$ with
  $\w\cup\atoms(\stk)\subseteq\atoms(\s)$ and
  $\emptyset\enty\stk\ofty\ty\FUNTY\ty'$, then
   \begin{align*}
     \config{\s}{\stk}{\UNBIND\,\val}\terminates
     &\;\bimp\; \config{\s}{\stk\comp(\vid.\,\UNBIND\,\vid)}{\val}\terminates
     &&\text{by definition of $\terminates$}\\
     &\;\bimp\; \config{\s}{\stk\comp(\vid.\,\UNBIND\,\vid)}{\val'}\terminates
     &&\text{since ${}\ent_{\w} \val\opeq\val'\ofty\ty\,\BINDTY$}\\
     &\;\bimp\; \config{\s}{\stk}{\UNBIND\,\val'}\terminates
     &&\text{by definition of $\terminates$}
   \end{align*}
   and thus ${}\ent_{\w} \UNBIND\,\val \opeq \UNBIND\,\val'\ofty
   \ATM\PRODTY\ty$.
   
   The proof of part (iii) is similar to that for (ii), using the
   frame $(\vid.\,\vid\,\val_1)$ in place of $(\vid.\,\UNBIND\,\vid)$.
\qed

\begin{lemma}
  \label{lem:howe-key}
  For all $n\geq 0$ and all $\w,\stk,\stk',\ty,\ty',\exp,\exp',\s$
  \begin{multline}
    \label{eq:5}
    \emptyset\ent_{\w} \stk \CR{\opeq^*} \stk' \ofty \ty\FUNTY\ty'
    \;\conj\; \emptyset\ent_{\w}\exp\opeq^*\exp'\ofty\ty \;\conj\;
    \atoms(\s)=\w \;\conj\;
    \config{\s}{\stk}{\exp}\terminates[n]\\
    \;\imp\; \config{\s}{\stk'}{\exp'}\terminates\;.
  \end{multline}
\end{lemma}
\proof
  The lemma is proved by induction on $n$. The base case $n=0$ follows
  from the definition of $\CR{-}$ (which implies that
  $\emptyset\ent_{\w}\es\CR{\opeq^*}\stk'\ofty\ty\FUNTY\ty'$ can only
  hold when $\stk'=\es$), combined with Lemma~\ref{lem:howe-1}(v) and
  the definition of $\opeqo$. For the induction step, assume
  \eqref{eq:5} holds and that
  \begin{align}
    &\emptyset\ent_{\w} \stk\CR{\opeq^*}\stk'\ofty\ty\FUNTY\ty'\label{eq:26}\\
    &\emptyset\ent_{\w} \exp \opeq^* \exp'\ofty\ty\label{eq:27}\\
    &\atoms(\s) = \w\label{eq:25}\\
    &\config{\s}{\stk}{\exp}\trans\config{\s_1}{\stk_1}{\exp_1}\label{eq:28}\\
    &\config{\s_1}{\stk_1}{\exp_1}\terminates^n\label{eq:29} 
  \end{align}
  We have to prove $\config{\s}{\stk'}{\exp'}\terminates$ and do so by
  an analysis of \eqref{eq:28} against the possible cases
  \ref{item:1}--\ref{item:9} in the definition
  of the transition relation in Figure~\ref{fig:trar}.

  \subsection*{Case \ref{item:1}.} In this case
  $\stk=\stk_1\comp(\vid.\,\exp_2)$, $\exp=\val\in\Val$, $\s_1=\s$, and
  $\exp_1=\exp_2\sub{\vid}{\val}$, for some $\exp_2$ and $\val$. For
  \eqref{eq:26} to hold, by definition of $\CR{\opeq^*}$ it must be the
  case that $\stk'=\stk_1'\comp(\vid.\,\exp_2')$ for some $\stk_1'$ and
  $\exp_2'$ with
  \begin{align}
    \{\vid\ofty\ty\} &\ent_{\w} \exp_2\opeq^* \exp_2'\ofty\ty_2\label{eq:30}\\
    \emptyset &\ent_{\w}
    \stk_1\CR{\opeq^*}\stk_1'\ofty\ty_2\FUNTY\ty'\label{eq:31}
  \end{align}
  for some type $\ty_2$. Since $\exp=\val$ is a value, applying
  Lemma~\ref{lem:howe-1}(v) to \eqref{eq:27} we get
  \begin{align}
    \emptyset &\ent_{\w} \val \opeq^* \val'\ofty\ty\label{eq:32}\\
    {} &\ent_{\w} \val'\opeq \exp'\ofty\ty\label{eq:33}
  \end{align}
  for some $\val'\in\Val$. Since $\opeq^*$ is substitutive
  (Lemma~\ref{lem:howe-1}(ii)), from \eqref{eq:30} and \eqref{eq:32}
  we get
  \begin{equation}
    \label{eq:34}
    \emptyset\ent_{\w} \exp_2\sub{\vid}{\val} \opeq^* 
    \exp_2'\sub{\vid}{\val'}\ofty \ty_2\;.
  \end{equation}
  Applying the induction hypothesis \eqref{eq:5} to \eqref{eq:31},
  \eqref{eq:34}, \eqref{eq:25} and to \eqref{eq:29}, we get
  $\config{\s}{\stk_1'}{\exp_2'\sub{\vid}{\val'}}\terminates$; hence
  $\config{\s}{\stk_1'\comp(\vid.\exp_2')}{\val'}\terminates$, that
  is, $\config{\s}{\stk'}{\val'}\terminates$; and therefore by
  \eqref{eq:33} we also have $\config{\s}{\stk'}{\exp'}\terminates$,
  as required.

  \subsection*{Case \ref{item:2}.} In this case we have $\exp = \LET\
  {\vid\EQ \exp_1} \IN \exp_2$, $\s_1=\s$ and
  $\stk_1=\stk\comp(\vid.\,\exp_2)$ for some $\exp_2$. Since
  \eqref{eq:27} holds, by definition of $\opeq^*$, there must exist
  some $\exp_1'$, $\exp_2'$ and $\ty_1$ with
  \begin{align}
    \emptyset &\ent_{\w} \exp_1 \opeq^*
    \exp_1'\ofty\ty_1\label{eq:35}\\
    \{\vid\ofty\ty_1\} &\ent_{\w} \exp_2 \opeq^*
    \exp_2'\ofty \ty\label{eq:36}\\
    {} &\ent_{\w} (\LET\ {\vid\EQ \exp_1'} \IN \exp_2')\opeq
    \exp'\ofty\ty\label{eq:37}\\
    \intertext{and then from \eqref{eq:26} and \eqref{eq:36} we get}
    \emptyset &\ent_{\w} \stk\comp(\vid.\,\exp_2) \CR{\opeq^*}
    \stk'\comp(\vid.\,\exp_2') \ofty \ty_1\FUNTY\ty'\;.\label{eq:38}
  \end{align}
  The induction hypothesis \eqref{eq:5} applied to \eqref{eq:38},
  \eqref{eq:35} and \eqref{eq:25} gives
  $\config{\s}{\stk'\comp(\vid.\,\exp_2')}{\exp_1'}\terminates$ and
  hence $\config{\s}{\stk'}{\LET\ {\vid\EQ \exp_1'} \IN
  \exp_2'}\terminates$. This and \eqref{eq:37} then give
  $\config{\s}{\stk'}{\exp'}\terminates$, as required.

  \subsection*{Case \ref{item:3}.}  This follows from the definition of
  $\opeq^*$ using its substitutivity property, much as for
  case~\ref{item:1}.

  \subsection*{Case \ref{item:4}.} In this case $\ty=\ty_1\PRODTY\ty_2$,
  $\exp=\PAIR{\val_1}{\val_2}$, $\s_1=\s$ and $\exp_1=\val_1$, for
  some $\ty_1,\ty_2\in\Ty$ and $\val_1,\val_2\in\Val$. By definition
  of $\CR{\opeq^*}$, for \eqref{eq:27} to hold it must be the case
  that
  \begin{align}
    \emptyset &\ent_{\w} \val_i\opeq^* 
    \val_i'\ofty\ty_i\quad\text{(for $i=1,2$)} \label{eq:39}\\
    {} &\ent_{\w} \PAIR{\val_1'}{\val_2'}\opeq
    \exp'\ofty\ty_1\PRODTY\ty_2 \label{eq:40}
  \end{align}
  for some $\val_1'$ and $\val_2'$.  By the induction hypothesis
  \eqref{eq:5} applied to \eqref{eq:26}, \eqref{eq:39}, \eqref{eq:25}
  and \eqref{eq:29}, we get $\config{\s}{\stk'}{\val_1'}\terminates$
  and hence also
  $\config{\s}{\stk'}{\FST\PAIR{v_1'}{v_2'}}\terminates$. Hence by
  \eqref{eq:40} we have $\config{\s}{\stk'}{\exp'}\terminates$, as
  required.

  \subsection*{Case \ref{item:5}.} This is similar to the previous
  case.

  \subsection*{Case \ref{item:6}.} In this case $\exp=\val_1\,\val_2$,
  $\s_1=\s$, $\stk_1=\stk$ and
  $\exp_1=\exp_2\sub{\vid[f],\vid}{\val_1,\val_2}$ for some
  $\val_1=\FUN(\vid[f]\,\vid\EQ \exp_2)$ and $\val_2$. Since
  \eqref{eq:27} holds, by definition of $\opeq^*$ together with
  Lemma~\ref{lem:howe-4}(iii), there must exist some $\exp_2'$,
  $\val_2'$ and $\ty_1$ with
  \begin{align}
    \{\vid[f]\ofty\ty_1\FUNTY\ty, \vid\ofty\ty_1\} &\ent_{\w}
    \exp_2\opeq^* \exp_2'\ofty\ty \label{eq:41}\\
    \emptyset &\ent_{\w} \val_2\opeq^* \val_2'\ofty\ty_1 \label{eq:42}\\
    {} &\ent_{\w} \FUN(\vid[f]\,\vid\EQ \exp_2')\,\val_2' \opeq
    \exp'\ofty\ty_1\FUNTY\ty\;. \label{eq:43}
  \end{align}
  Since $\opeq^*$ is compatible (Lemma~\ref{lem:howe-1}(ii)), from
  \eqref{eq:41} we get $\emptyset\ent_{\w} \val_1\opeq^*
  \FUN(\vid[f]\,\vid\EQ \exp_2') \ofty \ty_1\FUNTY\ty$; and since
  $\opeq^*$ is also substitutive, this together with \eqref{eq:41} and
  \eqref{eq:42} gives $\emptyset\ent_{\w}
  \exp_2\sub{\vid[f],\vid}{\val_1,\val_2} \opeq^*
  \exp_2'\sub{\vid[f],\vid}{\FUN(\vid[f]\,\vid\EQ \exp_2'), \val_2'}
  \ofty \ty$. Therefore by the induction hypothesis \eqref{eq:5}
  applied to \eqref{eq:26}, this, \eqref{eq:25} and \eqref{eq:29}, we
  get
  $\config{\s}{\stk'}{\exp_2'\sub{\vid[f],\vid}{\FUN(\vid[f]\,\vid\EQ
      \exp_2'), \val_2'}}\terminates$.  Hence
  $\config{\s}{\stk'}{\FUN(\vid[f]\,\vid \EQ
    \exp_2')\,\val_2'}\terminates$ and thus by \eqref{eq:43},
  $\config{\s}{\stk'}{\exp'}\terminates$ as required.

  \subsection*{Case \ref{item:7}.} In this case $\ty=\ATM$,
  $\exp=\FRESH\UNITVAL$, $\s_1=\s\ords\atm$, $\stk_1=\stk$ and
  $\exp_1=\atm$, for some $\atm\notin\atoms(\s)=\w$. Since
  \eqref{eq:27} holds, by definition of $\opeq^*$ we have
  \begin{equation}
   \label{eq:44}
    {}\ent_{\w} \FRESH\UNITVAL\opeq \exp'\ofty\ATM\;.
  \end{equation}
  By Lemma~\ref{lem:howe-2}(iii) applied to \eqref{eq:26}, we have
  $\emptyset\ent_{\w\cup \{\atm\}}\stk\CR{\opeq^*}
  \stk'\ofty\ATM\FUNTY\ty'$; and by Lemma~\ref{lem:howe-1}(iii) we
  also have $\emptyset\ent_{\w\cup\{\atm\}} \atm\opeq^*
  \atm\ofty\ATM$. So by the induction hypothesis \eqref{eq:5} applied
  to these, $\atoms(\s\ords\atm)=\w\cup\{\atm\}$ and \eqref{eq:29}, we
  get $\config{\s\ords a}{\stk'}{\atm}\terminates$.  Hence
  $\config{\s}{\stk'}{\FRESH}\terminates$ and hence from \eqref{eq:44}
  we also have $\config{\s}{\stk'}{e'}\terminates$, as required.

  \subsection*{Case \ref{item:8}.} In this case $\ty=\ATM\PRODTY\ty_1$,
  $\exp=\UNBIND\,\BINDVAL{\atm}{\val}$, $\s_1=\s\ords a_1$,
  $\stk_1=\stk$, and
  $\exp_1=\PAIR{\atm_1}{\val\rename{\atm}{\atm_1}}$, for some $\ty_1$,
  $\atm$, $\val$ and $\atm_1$ with $\atm_1\notin\atoms(\s)=\w$. Since
  \eqref{eq:27} holds, by definition of $\opeq^*$ together with parts
  (i) and (ii) of Lemma~\ref{lem:howe-4}, there must exist some
  $\val'$ with
  \begin{align}
    \emptyset &\ent_{\w} \val \opeq^* \val'\ofty\ty_1 \label{eq:45}\\
    {} &\ent_{\w} \UNBIND\,\BINDVAL{\atm}{\val'}\opeq 
    \exp' \ofty \ATM\PRODTY\ty_1\;. \label{eq:46}
  \end{align}
  We now appeal to the easily verified fact that since
  $\atm_1\notin\w\supseteq\atoms(\val,\val')$, the renamed values
  $\val\rename{\atm}{\atm_1}$ and $\val'\rename{\atm}{\atm_1}$ are
  respectively equal to the permuted values
  $\swap{\atm}{\atm_1}\act\val$ and $\swap{\atm}{\atm_1}\act \val'$
  (where $\swap{\atm}{\atm_1}$ denotes the permutation swapping $\atm$
  and $\atm'$).  Therefore by parts (i) and (ii) of
  Lemma~\ref{lem:howe-2} applied to \eqref{eq:45} and by parts (ii)
  and (iii) of Lemma~\ref{lem:howe-1}, we have
  \begin{equation}
    \label{eq:79}
    \emptyset\ent_{\w\cup\{\atm_1\}} 
    \PAIR{\atm_1}{\val\rename{\atm}{\atm_1}} \opeq^*
    \PAIR{\atm_1}{\val'\rename{\atm}{\atm_1}} \ofty\ATM\PRODTY\ty_1\;.
  \end{equation}
  By applying Lemma~\ref{lem:howe-2}(iii) to \eqref{eq:26} we also
  have
  \[
  \emptyset\ent_{\w\cup\{\atm_1\}} \stk\CR{\opeq^*}\stk'
  \ofty\ATM\PRODTY\ty_1\FUNTY\ty'\;.
  \]
  Then applying the induction hypothesis \eqref{eq:5} to this,
  \eqref{eq:79}, $\atoms(\s\ords\atm_1) = \w\cup\{\atm_1\}$ and
  \eqref{eq:29} yields $\config{\s\ords\atm_1}{\stk'}{\PAIR{\atm_1}
    {\val'\rename{\atm}{\atm_1}}}\terminates$.  Therefore
  $\config{\s}{\stk'}{\UNBIND\,\BINDVAL{\atm}{\val'}}\terminates$; and
  hence by \eqref{eq:46}, we also have
  $\config{\s}{\stk'}{e'}\terminates$, as required.

  \subsection*{Case \ref{item:9}.} In this case $\ty=\NAT$,
  $\exp=\OBS\,\atm_1\ldots\atm_k$ for some
  $\atm_1,\ldots,\atm_k\in\w$, $\s_1=\s$, $\stk_1=\stk$, and
  $\exp_1=\rep{m}$ where $m=\den{\OBS}_{\s}(\atm_1,\ldots,\atm_k)$.
  Since \eqref{eq:27} holds, by definition of $\opeq^*$ together with
  Lemma~\ref{lem:howe-4}(i), we must have
  \begin{equation}
    \label{eq:48}
    {}\ent_{\w} \OBS\,\atm_1\ldots\atm_k \opeq e'\ofty \NAT\;.
  \end{equation}
  Note that by Lemma~\ref{lem:howe-1}(iii) we also have
  $\emptyset\ent_{\w}\rep{m}\opeq^*\rep{m}\ofty\NAT$. So by the
  induction hypothesis \eqref{eq:5} applied to this, \eqref{eq:26},
  \eqref{eq:25} and \eqref{eq:29} we get
  $\config{\s}{\stk'}{\rep{m}}\terminates$. Since
  $m=\den{\OBS}_{\s}(\atm_1,\ldots,\atm_k)$, this implies that
  $\config{\s}{\stk'}{\OBS\,\atm_1\ldots\atm_k}\terminates$; and
  hence from \eqref{eq:48} we have that
  $\config{\s}{\stk'}{\exp'}\terminates$ holds, as required.

  \bigskip\noindent
  This completes the proof of Lemma~\ref{lem:howe-key}.
\qed

\begin{lemma}
  \label{lem:howe-5}
  Let $(\opeq^*)^{+}$ denote the transitive closure of $\opeq^*$.
  Then 
  \[
  \Gamma\ent_{\w}\exp\opeq^*\exp'\ofty\ty \;\imp\;
  \Gamma\ent_{\w}\exp' \mathrel{(\opeq^*)^{+}} \exp\ofty\ty\;.
  \]
\end{lemma}
\proof
  This can be proved by induction on the derivation of
  $\Gamma\ent_{\w}\exp\opeq^*\exp'\ofty\ty$ from the rule in
  \eqref{eq:4} and the rules for compatible refinement in
  Figure~\ref{fig:comr}, using the fact that $\opeqo$ is symmetric and
  using Lemmas~\ref{lem:howe-3} and \ref{lem:howe-1}(iii).
\qed

We can now complete the proof of Theorem~\ref{thm:ciu} by showing that
$\opeqo$ is compatible and substitutive (Definition~\ref{def:expr}).
Since $\opeq^*$ has those properties by Lemma~\ref{lem:howe-1}(ii), it
suffices to show that $\opeqo$ coincides with $\opeq^*$. We already
noted in Lemma~\ref{lem:howe-3} that $\opeqo$ is contained in
$\opeq^*$. For the reverse inclusion, since $\opeq^*$ is substitutive
and reflexive (Lemma~\ref{lem:howe-1}), it is closed under
substituting values for variables; so by Definition~\ref{def:opeqo}, it
suffices to show that
\begin{equation}
  \label{eq:47}
  \emptyset\ent_{\w} \exp\opeq^*\exp' \ofty\ty \;\imp\; 
  {}\ent_{\w} \exp\opeq \exp'\ofty\ty\;.
\end{equation}
To see this, note that by Lemma~\ref{lem:howe-key} (together with
Lemmas~\ref{lem:howe-1}(iv) and \ref{lem:howe-2}(ii)) we have:
\begin{multline}
  \label{eq:49}
  \emptyset\ent_{\w} \exp\opeq^* \exp' \ofty\ty \;\imp\; \forall
  \s,\stk,\ty'.\; \w\cup\atoms(\stk)\subseteq\atoms(\s) \;\conj\;
  \emptyset\enty\stk\ofty\ty\FUNTY\ty' \;\conj\;
  \config{\s}{\stk}{\exp}\terminates \\
  \;\imp\; \config{\s}{\stk}{\exp'}\terminates\;.
\end{multline}
Since the right-hand side of the implication in \eqref{eq:49} is a
transitive relation between expressions $\exp,\exp'$, we also have
\begin{multline*}
  \emptyset\ent_{\w} \exp\mathrel{{\opeq^*}^+} \exp' \ofty\ty \;\imp\;
  \forall \s,\stk,\ty'.\; \w\cup\atoms(\stk)\subseteq\atoms(\s) \;\conj\;
  \emptyset\enty\stk\ofty\ty\FUNTY\ty' \;\conj\;
  \config{\s}{\stk}{\exp}\terminates\\
  \;\imp\; \config{\s}{\stk}{\exp'}\terminates
\end{multline*}
and therefore Lemma~\ref{lem:howe-5} gives
\begin{multline}
  \label{eq:50}
  \{\}\ent_{\w} \exp\opeq^* \exp' \ofty\ty \;\imp\; \forall
  \s,\stk,\ty'.\; (\w\cup\atoms(\stk)\subseteq\atoms(\s) \;\conj\;
  \emptyset\enty\stk\ofty\ty\FUNTY\ty' \;\conj\;
  \config{\s}{\stk}{\exp'}\terminates\\
  \;\imp\; \config{\s}{\stk}{\exp}\terminates\;.
\end{multline}
Combining \eqref{eq:49} and \eqref{eq:50} gives \eqref{eq:47}.  \qed

\section{Proof of Proposition~\ref{prop:ext-bind-1}}
\label{app:proof-ext-bind-1-proposition}

Let ${\er}$ be the closure under compatible refinement
(Figure~\ref{fig:comr}) of the pairs of closed atom binding values
that we wish to show are operationally equivalent. In other words
$\er$ is the expression relation inductively defined by the following
two rules.
\begin{equation}
  \label{eq:8}
  \inferrule{%
    \atm''\notin\w \subseteq\atoms(\atm,\val,\atm',\val')\\
    {}\ent_{\w\cup\{\atm''\}} \val\rename{\atm}{\atm''} \opeq 
    \val'\rename{\atm'}{\atm''} \ofty \ty
    }{%
    \emptyset\ent_{\w} \BINDVAL{\atm}{\val} \er \BINDVAL{\atm'}{\val'} \ofty
    \ty\,\BINDTY}
  \qquad
  \inferrule{%
    \Gamma\ent_{\w} \exp \CR{\er} \exp' \ofty \ty
  }{%
    \Gamma\ent_{\w} \exp \er \exp' \ofty \ty}    
\end{equation}

\begin{lemma}
  \label{lem:ext-bind-1}\hfill
  \begin{itemize}
  \item[(i)] $\er$ is compatible and substitutive.

  \item[(ii)] $\atoms(\exp)\subseteq\w \;\conj\;
    \Gamma\enty\exp\ofty\ty \;\imp\;
    \Gamma\ent_{\w}\exp\er\exp\ofty\ty$.

  \item[(iii)] $\atoms(\stk)\subseteq\w \;\conj\;
    \Gamma\enty\stk\ofty\ty\FUNTY\ty' \;\imp\; \Gamma\ent_{\w}\stk
    \CR{\er}\stk \ofty\ty\FUNTY\ty'$.

  \item[(iv)] $\Gamma\ent_{\w}\val\er\exp'\ofty\ty \;\imp\;
    \exp'\in\Val$.
  \end{itemize}
\end{lemma}
\proof
  These properties of $\er$ are simple consequences of its definition
  in \eqref{eq:8}, the definition of compatible refinement in
  Figure~\ref{fig:comr}, and the definition of its extension to a
  relation between frame stacks given by the last two rules in that
  figure.
\qed

\begin{lemma}
  \label{lem:ext-bind-2}\hfill
  \begin{itemize}
  \item[(i)] $\er$ is equivariant.

  \item[(ii)] $\Gamma\ent_{\w}\exp\er\exp'\ofty\ty \;\conj\;
    \w\subseteq\w' \;\imp\; \Gamma\ent_{\w'}\exp\er\exp'\ofty\ty$.

  \item[(iii)] $\Gamma\ent_{\w}\stk \CR{\er} \stk'
    \ofty\ty\FUNTY\ty' \;\conj\; \w\subseteq\w' \;\imp\;
    \Gamma\ent_{\w'}\stk \er\stk'\ofty\ty\FUNTY\ty'$.
  \end{itemize}
\end{lemma}
\proof
  This is the analogue of Lemma~\ref{lem:howe-2} for $\er$, and is
  proved in the same way as that lemma.
\qed

\begin{lemma}
  \label{lem:ext-bind-3}
  For all $n\geq 0$ and all $\w,\stk,\stk',\ty,\ty',\exp,\exp',\s$
  \begin{equation}
    \label{eq:51}
    \emptyset\ent_{\w} \stk \CR{\er} \stk' \ofty \ty\FUNTY\ty'
    \;\conj\; \emptyset\ent_{\w}\exp\er\exp'\ofty\ty \;\conj\;
    \atoms(\s)=\w \;\conj\;
    \config{\s}{\stk}{\exp}\terminates[n]
    \;\imp\; \config{\s}{\stk'}{\exp'}\terminates\;.
  \end{equation}
\end{lemma}
\proof
  The lemma is proved by induction on $n$. The base case $n=0$ follows
  directly from Lemma~\ref{lem:ext-bind-1}(iii) and the definition of
  $\CR{\er}$ (which implies that
  $\{\}\ent_{\w}\es\CR{\er}\stk'\ofty\ty\FUNTY\ty'$ can only hold when
  $\stk'=\es$). For the induction step, assume \eqref{eq:51} holds and
  that
  \begin{align}
    &\emptyset\ent_{\w} \stk\CR{\er}\stk'\ofty\ty\FUNTY\ty'\label{eq:53}\\
    &\emptyset\ent_{\w} \exp \er \exp'\ofty\ty\label{eq:54}\\
    &\atoms(\s) = \w\label{eq:52}\\
    &\config{\s}{\stk}{\exp}\trans\config{\s_1}{\stk_1}{\exp_1}\label{eq:55}\\
    &\config{\s_1}{\stk_1}{\exp_1}\terminates^n\label{eq:56}
  \end{align}
  We have to prove $\config{\s}{\stk'}{\exp'}\terminates$ and do so by
  an analysis of \eqref{eq:55} against the possible cases
  \ref{item:1}--\ref{item:9} in the definition of the transition
  relation in Figure~\ref{fig:trar}. Cases~\ref{item:1}, \ref{item:3}
  and \ref{item:6} follow from the definition of $\er$ and its
  substitutivity property; we give the details for the first one and
  omit the other two. Cases \ref{item:4}, \ref{item:5} and
  \ref{item:9} also follow easily from the definition of $\er$ (using
  Lemma~\ref{lem:ext-bind-1}(ii) in the third case). So we give the
  proofs just for cases \ref{item:1}, \ref{item:2}, \ref{item:7} and
  \ref{item:8}.

  \subsection*{Case \ref{item:1}.} In this case
  $\stk=\stk_1\comp(\vid.\,\exp_2)$, $\exp=\val\in\Val$, $\s_1=\s$, and
  $\exp_1=\exp_2\sub{\vid}{\val}$, for some $\exp_2$ and $\val$. For
  \eqref{eq:53} to hold, by definition of $\CR{\er}$ it must be the
  case that $\stk'=\stk_1'\comp(\vid.\,\exp_2')$ for some $\stk_1'$ and
  $\exp_2'$ with
  \begin{align}
    \{\vid\ofty\ty\} &\ent_{\w} \exp_2\er
    \exp_2'\ofty\ty_2 \label{eq:57}\\
    \emptyset &\ent_{\w}
    \stk_1\CR{\er}\stk_1'\ofty\ty_2\FUNTY\ty'\label{eq:58}
  \end{align}
  for some type $\ty_2$. Since $\exp=\val$ is a value, applying
  Lemma~\ref{lem:ext-bind-1}(iv) to \eqref{eq:54} we get
  $\exp'=\val'$ for some $\val'\in\Val$. So since $\CR{\er}$ is
  substitutive (Lemma~\ref{lem:ext-bind-1}(i)), from \eqref{eq:54} and
  \eqref{eq:57} we get
  \begin{equation}
    \label{eq:59}
    \emptyset\ent_{\w} \exp_2\sub{\vid}{\val} \er  
    \exp_2'\sub{\vid}{\val'}\ofty \ty_2\;.
  \end{equation}
  Applying the induction hypothesis \eqref{eq:51} to \eqref{eq:58},
  \eqref{eq:59}, \eqref{eq:52} and to \eqref{eq:56}, we get
  $\config{\s}{\stk_1'}{\exp_2'\sub{\vid}{\val'}}\terminates$; hence
  $\config{\s}{\stk_1'\comp(\vid.\exp_2')}{\val'}\terminates$,
  that is, $\config{\s}{\stk'}{\exp'}\terminates$, as required.

  \subsection*{Case \ref{item:2}.} In this case $\exp = \LET\ {\vid\EQ
    \exp_1} \IN \exp_2$, $\s_1=\s$ and
  $\stk_1=\stk\comp(\vid.\,\exp_2)$ for some $\exp_2$.  For
  \eqref{eq:54} to hold, by definition of $\CR{\er}$ it must be the
  case that $\exp'=\LET\ {\vid\EQ\exp_1'} \IN \exp_2'$ for some
  $\exp_1',\exp_2'$ and $\ty_1$ with
  \begin{align}
    \{\} &\ent_{\w} \exp_1\er \exp_1'\ofty \ty_1\label{eq:60}\\
    \{x\ofty\ty_1\} &\ent_{\w} \exp_2 \er \exp_2'\ofty \ty\;.
    \label{eq:61}
  \end{align}
  From \eqref{eq:53} and \eqref{eq:61} we get $\emptyset\ent_{\w}
  \stk\comp(\vid.\,\exp_2) \CR{\er} \stk'\comp(\vid.\,\exp_2')\ofty
  \ty\FUNTY \ty'$; and the induction hypothesis \eqref{eq:51} applied
  to this, \eqref{eq:60}, \eqref{eq:52} and \eqref{eq:56} gives
  $\config{\s}{\stk'\comp(\vid.\,\exp_2')}{\exp_1'}\terminates$. Hence
  $\config{\s}{\stk'}{\LET\ {\vid\EQ\exp_1'} \IN \exp_2'}\terminates$,
  that is, $\config{\s}{\stk'}{\exp'}\terminates$, as required.

  \subsection*{Case \ref{item:7}.} In this case $\ty=\ATM$,
  $\exp=\FRESH\UNITVAL$, $\s_1=\s\ords\atm$, $\stk_1=\stk$ and
  $\exp_1=\atm$, for some atom $\atm\notin\w$. For \eqref{eq:54} to
  hold, by definition of ${\er}$ it must be the case that
  $\exp'=\FRESH\UNITVAL$.  Now Lemma~\ref{lem:ext-bind-2}(iii) applied
  to \eqref{eq:53} gives $\emptyset\ent_{\w\cup\{\atm\}}\stk\CR{\er}
  \stk'\ofty\ty\FUNTY\ty'$; and Lemma~\ref{lem:ext-bind-1}(ii) gives
  $\emptyset\ent_{\w\cup\{\atm\}} \atm\er \atm\ofty \ATM$.  Applying
  the induction hypothesis \eqref{eq:51} to these two facts,
  $\atoms(\s\ords\atm) = \w\cup\{\atm\}$ and \eqref{eq:56} gives
  $\config{\s\ords\atm}{\stk'}{\atm}\terminates$. Hence
  $\config{\s}{\stk'}{\FRESH\UNITVAL}\terminates$,
  that is, $\config{\s}{\stk'}{\exp'}\terminates$, as required.

  \subsection*{Case \ref{item:8}.} In this case $\ty=\ATM\PRODTY\ty_1$,
  $\exp=\UNBIND\,\BINDVAL{\atm}{\val}$, $\s_1=\s\ords \atm_1$,
  $\stk_1=\stk$, and
  $\exp_1=\PAIR{\atm_1}{\val\rename{\atm}{\atm_1}}$, for some $\ty_1$,
  $\atm$, $\val$ and $\atm_1$ with $\atm_1\notin\w$.  For
  \eqref{eq:54} to hold, by definition of $\er$ it must be the case
  that $e'=\UNBIND\,\BINDVAL{\atm'}{\val'}$ with
  \begin{equation}
   \label{eq:62}
    \begin{array}{rl}
      \text{either (a):} & \atm=\atm'\;\conj\;
      \emptyset\ent_{\w} \val\er \val'\ofty\ty_1\\
      \text{or (b):} & \exists \atm''\notin\w.\; 
      {}\ent_{\w\cup\{\atm''\}} \val\rename{\atm}{\atm''} \opeq 
      \val'\rename{\atm}{\atm''} \ofty\ty_1
    \end{array}
  \end{equation}
  
  If \eqref{eq:62}(a) holds, then as in the proof of
  Lemma~\ref{lem:howe-key} we now appeal to the easily verified fact
  that since $\atm_1\notin\w\supseteq\atoms(\val,\val')$, the renamed
  values $\val\rename{\atm}{\atm_1}$ and $\val'\rename{\atm}{\atm_1}$
  are respectively equal to the permuted values
  $\swap{\atm}{\atm_1}\act\val$ and $\swap{\atm}{\atm_1}\act \val'$
  (where $\swap{\atm}{\atm_1}$ denotes the permutation swapping $\atm$
  and $\atm'$).  Therefore from the fact that $\emptyset\ent_{\w}
  \val\er \val'\ofty\ty_1$ holds, from parts (i) and (ii) of
  Lemma~\ref{lem:ext-bind-2} we get $\emptyset\ent_{\w\cup\{\atm_1\}}
  \val\rename{\atm}{\atm_1} \er \val'\rename{\atm}{\atm_1}
  \ofty\ty_1$.  Then since $\atm=\atm'$, by
  Lemma~\ref{lem:ext-bind-1}(ii) we have
  $\emptyset\ent_{\w\cup\{\atm_1\}}
  \PAIR{\atm_1}{\val\rename{\atm}{\atm_1}} \er
  \PAIR{\atm_1}{\val'\rename{\atm'}{\atm_1}} \ofty \ATM\PRODTY\ty_1$.
  Applying the induction hypothesis \eqref{eq:51} to this,
  \eqref{eq:53} (weakened using Lemma~\ref{lem:ext-bind-2}(iii)),
  $\atoms(\s\ords\atm_1) = \w\cup\{\atm_1\}$ and \eqref{eq:56} yields
  $\config{\s\ords
    \atm_1}{\stk'}{\PAIR{\atm_1}{\val'\rename{\atm}{\atm_1}}}\terminates$
  with $\atm_1\notin\atoms(\s)$.  Therefore by definition of
  $\terminates$, we also have
  $\config{\s}{\stk'}{\UNBIND\,\BINDVAL{\atm'}{\val'}}\terminates$.
  
    If \eqref{eq:62}(b) holds, then by Theorem~\ref{thm:ciu}, so does
  \begin{equation}
    \label{eq:63}
    {}\ent_{\w\cup\{\atm''\}} \PAIR{\atm''}{\val\rename{\atm}{\atm''}}
    \opeq
    \PAIR{\atm''}{\val'\rename{\atm}{\atm''}} \ofty\ATM\PRODTY\ty_1 
  \end{equation}
  Lemma~\ref{lem:ter-equivar} applied to \eqref{eq:56} with $\pi=
  \swap{\atm_1}{\atm''}$ gives $\config{\s\ords
    \atm''}{\stk}{\PAIR{\atm''}{\val\rename{\atm}{\atm''}}}\terminates[n]$.
  Combining this with \eqref{eq:53} (weakened using
  Lemma~\ref{lem:ext-bind-2}(iii)), $\emptyset\ent_{\w\cup\{\atm''\}}
  \PAIR{\atm''}{\val\rename{\atm}{\atm''}} \er
  \PAIR{\atm''}{\val\rename{\atm}{\atm''}} \ofty \ATM\PRODTY\ty_1$ (by
  Lemma~\ref{lem:ext-bind-1}(ii)),
  $\atoms(\s\ords\atm'')=\w\cup\{\atm''\}$ and the induction
  hypothesis \eqref{eq:51}, we get $\config{\s\ords
    \atm''}{\stk'}{\PAIR{\atm''}{\val\rename{\atm}{\atm''}}}\terminates$.
  Then by definition of $\opeq$, from this and \eqref{eq:63} we get
  $\config{\s\ords\atm''}{\stk'}{\PAIR{\atm''}
    {\val'\rename{\atm}{\atm''}}}\terminates$ with $\atm''\notin\s$.
  Therefore as before, by definition of $\terminates$, we also have
  $\config{\s}{\stk'}{\UNBIND\,\BINDVAL{\atm'}{\val'}}\terminates$.

  So in either case in \eqref{eq:62}, since
  $\exp'=\UNBIND\,\BINDVAL{\atm'}{\val'}$, we get
  $\config{\s}{\stk'}{\exp'}\terminates$, as required.

    \bigskip\noindent 
    This completes the proof of Lemma~\ref{lem:ext-bind-3}.
\qed

We can now complete the proof of Proposition~\ref{prop:ext-bind-1}.
For any type $\ty\in\Ty$, suppose we are given closed, well-typed atom
binding values $\emptyset\enty\BINDVAL{\atm}{\val}\ofty\ty\,\BINDTY$
and $\emptyset\enty\BINDVAL{\atm'}{\val'}\ofty\ty\,\BINDTY$ with
$\atoms(\atm,\val,\atm',\val')\subseteq\w$ and satisfying
\begin{equation}
  \label{eq:67}
  {}\ent_{\w\cup\{\atm''\}}\val\rename{\atm}{\atm''} \opeq
  \val'\rename{\atm'}{\atm''} \ofty \ty
\end{equation}
for some atom $\atm''\notin\w$. By definition of $\er$ this implies
\begin{equation}
  \label{eq:65}
  \emptyset\ent_{\w} \BINDVAL{\atm}{\val} \er \BINDVAL{\atm'}{\val'}
  \ofty \ty\,\BINDTY\;. 
\end{equation}
For any $\w'$, $\s$, $\stk$, and $\ty'$ with
$\atoms(\s)=\w'\supseteq\w\cup\atoms(\stk) $ and
$\emptyset\enty\stk\ofty\ty\FUNTY\ty'$, we have
\begin{equation}
  \label{eq:66}
  \emptyset\ent_{\w'} \stk\CR{\er} \stk\ofty
\ty\FUNTY\ty'
\end{equation}
by Lemma~\ref{lem:ext-bind-1}(iii) and
\begin{equation}
  \label{eq:68}
  \emptyset\ent_{\w'} \BINDVAL{\atm}{\val} \er \BINDVAL{\atm'}{\val'}
  \ofty \ty\,\BINDTY
\end{equation}
by Lemma~\ref{lem:ext-bind-2}(ii) applied to \eqref{eq:65}.
So Lemma~\ref{lem:ext-bind-3}
applied to \eqref{eq:66}, \eqref{eq:68} and $\atoms(\s)=\w'$, we have
\[
\config{\s}{\stk}{\BINDVAL{\atm}{\val}}\terminates
\;\imp\;
\config{\s}{\stk}{\BINDVAL{\atm'}{\val'}}\terminates\;.
\]
Since $\opeq$ is symmetric, the same argument shows that \eqref{eq:67}
implies
\[
\config{\s}{\stk}{\BINDVAL{\atm'}{\val'}}\terminates \;\imp\;
\config{\s}{\stk}{\BINDVAL{\atm}{\val}}\terminates\;.
\]
Thus \eqref{eq:67} implies that $\BINDVAL{\atm}{\val}$ and
$\BINDVAL{\atm'}{\val'}$ are operationally equivalent, as required.
\qed

\end{document}